\documentclass[journal]{IEEEtran}
\IEEEoverridecommandlockouts
\usepackage{cite}
\usepackage{amsmath,amssymb,amsfonts}
\usepackage{algorithmic}
\usepackage{graphicx}
\usepackage{textcomp}
\usepackage[table]{xcolor}
\usepackage{siunitx}
\usepackage[T1]{fontenc}
\usepackage{subcaption} % replaces \usepackage{subfigure}
\usepackage{comment}
\usepackage{booktabs,tabularx}
\usepackage{lipsum}

\usepackage{color,soul}

\def\BibTeX{{\rm B\kern-.05em{\sc i\kern-.025em b}\kern-.08em
    T\kern-.1667em\lower.7ex\hbox{E}\kern-.125emX}}

\begin{document}

\title{Digital Twin Enabled Talkative Power Conversion}

\author{Peter A. Hoeher, Yang Leng, Maximilian Mewis-Noll, and Rongwu Zhu}

\maketitle

\begin{abstract}
Talkative Power Conversion (TPC) is a paradigm in which communication is integrated directly into switched-mode power electronic converters (PECs) by modulating their inherent switching ripple. 
TPC enables simultaneous power transfer and information exchange without dedicated transmitter-side communication hardware, and avoids additional wiring. 
This paper combines, for the first time, the Digital Twin (DT) concept with the TPC paradigm. 
By establishing interaction between one or several physical power converters and their virtual counterpart(s), the DT framework enables enhanced modeling, prediction, and optimization of individual or networked PECs. 
We distinguish three coupling architectures--embedded DT, remote DT, and networked DT--and demonstrate the added value of combining DT and TPC concepts when the DT and PEC are not co-located. 
The proposed framework is based on state-space modeling in conjunction with Kalman filtering and facilitates joint treatment of data transmission, parameter estimation, and system operation, thereby improving robustness, adaptability, and power efficiency.
The presented approach highlights the potential of DT-enhanced TPC as a key enabler for future cyber-physical energy systems with tightly integrated power and communication functionalities.
\end{abstract}


\markboth{}{}

\section{Introduction}\label{sec:Introduction}
The evolution of power electronics (PE) is increasingly marked by an integration of computation and digital communication, referred to as computational power electronics (CPE) \cite{Beiranvand2025}. 
CPE technologies are expected to pave the road to high-reliability PE \cite{Alcaide2024}. 
A number of computational technologies have recently emerged in the field of PE, including edge and edge-cloud computing assisted PE systems \cite{Cao2021}, cyber-physical power systems (CPPS) \cite{Yohanandhan2020, Mazumder2021}, artificial intelligence (AI) and machine-learning (ML) \cite{Bose2020, Chen2025}, and digital twins (DTs) \cite{Chen2023a, Bai2024, Wu2025, Bhoi2025}. 
A DT is a dynamic virtual replica of a physical system that is continuously updated with real-time sensor data. 
DT technology is an emerging area, and has extensively been studied for industrial applications \cite{Tao2018, Tao2024}. 
This paper focuses on DTs applied to power electronic converters (PECs)--a highly active research area, see for example \cite{Jain2020, Milton2020, Peng2021, Zhao2022, Lei2023, DiNezio2023, deLopezDiz2023, Song2024, Kuprat2024, Liu2024, BehzadNazif2025, Zheng2025, Wang2026b, Blaabjerg2026} (in chronological order). 
Potential tasks include diagnostics \cite{Jain2020, Milton2020}, condition monitoring \cite{deLopezDiz2023, Wang2026b}, health indication \cite{Peng2021, DiNezio2023}, control and predictive control \cite{Lei2023, Song2024, Zheng2025}, parameter identification and degradation monitoring \cite{Zhao2022, Liu2024}, thermal monitoring \cite{Kuprat2024}, and reliability predictions \cite{Blaabjerg2026}. 

\begin{figure}[t]
\centering
\includegraphics[width=0.8\linewidth]{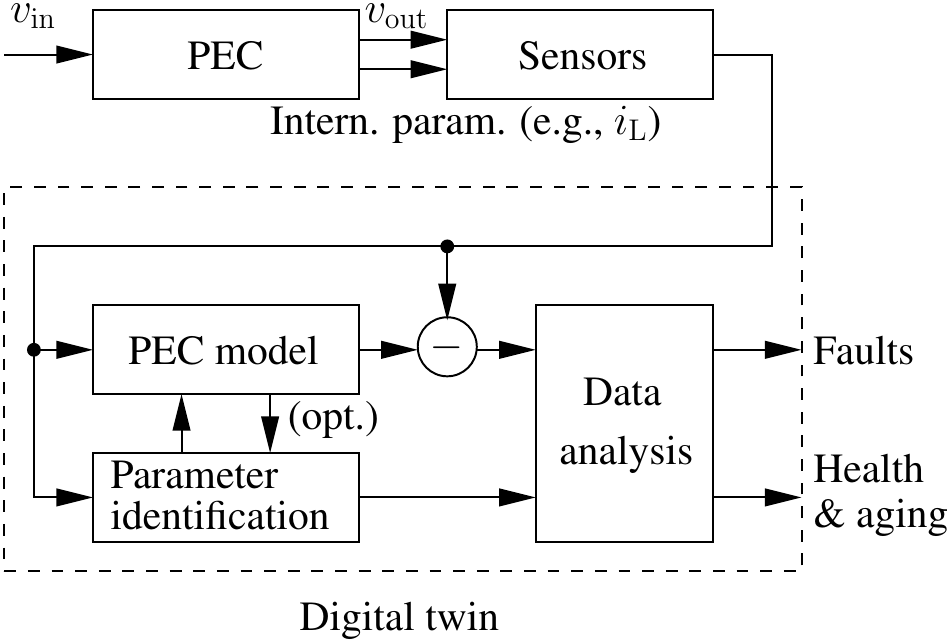}
\caption{Schematic of the digital twin concept applied to a PEC.} 
\label{fig:pec-dt}
\end{figure}

A DT, called cyber system or digital system, is a virtual replica of a physical system, operating in parallel with its physical counterpart (see Fig.~\ref{fig:pec-dt}). 
With a sufficient number of sensor signals at adequate fidelity, the DT can accurately predict relevant system parameters. 
Coupling a DT with a PEC or a PEC-integrated power grid can exhibit different levels of interaction, synchronization, and intelligence. 
In TABLE~\ref{coupling}, fundamental PEC-to-DT coupling architectures reported in the literature (see e.\,g. \cite{Chen2023a, Bai2024, Wu2025, Bhoi2025}) are classified, ranging from simple monitoring to highly integrated intelligent DTs.
Important design metrics include model fidelity, information transmission, adaptivity, and latency.

\begin{table*}[t]
\caption{Classification of fundamental PEC-to-DT coupling architectures, ranging from simple monitoring to highly integrated intelligent DTs.}
\centering
\begin{tabularx}{\textwidth}{X}
%%\begin{tabular}{l}
\toprule
%\begin{itemize}
%\item 
{\bf One-way coupling (physical to cyber domain):} 
In monitoring or shadow mode, sensors measure quantities such as voltages, currents, temperatures, and switching states, and stream this data into the DT. 
The DT reconstructs internal variables (e.\,g., losses), performs condition monitoring, and estimates quantities that are difficult to measure directly. 
Typical applications include fault detection, thermal prediction, and efficiency tracking \cite{Wu2025}.\\
%\item 
\midrule
{\bf Two-way coupling (closed-loop interaction):} 
In DT-assisted control, the controller uses predictions generated by the DT, such as current ripple, to adapt switching frequency, modulation strategy, or other control parameters. 
Common examples include model predictive control (MPC) and online parameter tuning. 
In this case, the DT becomes an integral part of the control loop \cite{Lei2023, Zheng2025}.\\
%\item 
\midrule
{\bf Real-time co-simulation (synchronous coupling):} 
Here, the DT runs in real time alongside the physical converter, often implemented on field-programmable gate array (FPGA) or real-time simulation platforms \cite{Milton2020}. 
Data exchange occurs at very fast time scales (e.\,g., several times per switching period), enabling cycle-accurate predictions and realistic emulation of internal states. 
This approach allows controllers to be tested under highly realistic operating conditions and provides tightly synchronized interaction between physical and virtual systems \cite{Zheng2025}.\\
%\item 
\midrule
{\bf Hardware-in-the-loop (HIL) coupling:} 
In controller HIL (CHIL), the controller hardware is real while the converter is simulated. 
In power HIL (PHIL), real power hardware interacts with a simulated environment. 
These configurations are widely used for controller validation, testing of fault scenarios, and verification prior to deployment. 
When integrated with a DT, the simulated environment provides a flexible and realistic testbed \cite{Milton2020, Barreto2025}.\\
%\item 
\midrule
{\bf Data-driven/AI-enhanced coupling:} 
A learning-based DT uses both historical and real-time data to continuously update its models. 
It can capture effects such as component aging (e.\,g., capacitor degradation) and nonlinear behaviors that are difficult to model analytically. 
This enables advanced applications like predictive maintenance, lifetime estimation, and fault prognosis, where the DT combines physical measurements with continuously learned data-driven models \cite{Blaabjerg2026}.\\
%\item 
\midrule
{\bf Multi-physics coupling:}
In this case, the DT incorporates multiple physical domains, including electrical phenomena (e.\,g., switching), thermal dynamics (e.\,g., junction temperature) \cite{Kuprat2024}, and mechanical effects (e.\,g., vibration). 
This type of coupling is especially important in high-power and reliability-critical systems \cite{Notash2026}.\\
%\item 
\midrule
{\bf Hierarchical/system-level coupling:} 
The converter DT is embedded within a larger system-level DT, such as those representing smart grids, electric vehicles, or renewable energy systems. 
Interaction occurs on multiple levels, from local switching and control to system-wide aspects like power transfer and grid stability. 
This hierarchical integration enables coordinated and system-wide optimization \cite{Bhoi2025}.\\
%\item 
\midrule
{\bf Edge vs.~cloud coupling:} 
The deployment location of the DT significantly affects its role. 
Edge-based DTs execute locally and provide fast responses suitable for real-time control, while cloud-based DTs handle computationally intensive tasks such as large-scale analytics. 
Hybrid approaches combine both, with fast local control loops and slower cloud-based optimization processes \cite{Cao2021, Bhoi2025}.\\
%\end{itemize}
\bottomrule
%%\end{tabular}
\end{tabularx}
\label{coupling}
\end{table*}

A key feature of DT technology is the continuous data transmission between the physical and digital domains, enabled by an external communication network and suitable protocols. 
This data transmission is related to the communication in modern distributed PE systems, which require tight coordination among parallel converters, distributed sources, and system-level functions such as power sharing, loss reduction, and protection. 
Conventional solutions rely on external communication infrastructures, e.\,g., power-line communication (PLC), fieldbus systems, or wireless links \cite{Hoeher2026b}, which introduce additional hardware complexity, cost, and potential electromagnetic interference (EMI) and reliability issues.

Another key feature of DT technology is the tracking algorithm, which determines the adaptivity of the DT to parameter changes of the physical asset.
Various tracking methods have been studied, including: 
(i) Milton {\em et al.} \cite{Milton2020} proposed an approach for the online diagnostic analysis of PECs utilizing real-time, probabilistic digital twinning. A DT of a PEC is treated as a real-time, probabilistic simulation model with random variables, developed using generalized polynomial chaos expansion (PCE). A DT model is partitioned in perspective of PEC control layers. Real-time solvers of the probabilistic DT model are embedded into PEC controllers running on FPGA devices. 
(ii) Peng {\em et al.} \cite{Peng2021} proposed a noninvasive health indicator estimation method based on the DT concept aiming for condition monitoring of PECs. Particle swarm optimization (PSO) is applied to estimate the unknown circuit parameters of interest based on the incoming data from both the DT and the physical prototype. 
(iii) Zhao {\em et al.} \cite{Zhao2022} proposed a physics-informed machine learning (PIML) parameter estimation method demonstrated for a buck converter. Condition monitoring is suggested as a possible task, among others.  A deep neural network and dynamic models of the converter
are seamlessly coupled to overcome the challenges related to training data, accuracy, and robustness. 
(iv) Lei {\em et al.} \cite{Lei2023} applied DT technology to the monitoring and control of DC--DC converters. Parameter identification is performed by Kalman filtering (KF). Four scenarios are investigated: adaptation to output voltage variations, adaptation to input voltage variations, investigation of controller failure, and switch replacement. 
(v) DiNezio {\em et al.} \cite{DiNezio2023} enabled real-time monitoring of degradation parameters of a DC--DC boost converter by using measurement data from the asset’s existing sensors, providing the necessary information for its monitoring and control. The data analysis is done by applying heuristic algorithms like PSO.
(vi) De Lopez Diz {\em et al.} \cite{deLopezDiz2023} described a DT concept applied to condition monitoring of a three-layer neutral point clamped (3L-NPC) converter. The combination of PSO and a genetic algorithm (GA) is employed to estimate the circuit parameters of interest, specifically the degradation of the output LC filter.
(vii) Song {\em et al.} \cite{Song2024} proposed a DT modeling method of a three-phase inverter-driven permanent magnet synchronous motor (PMSM) for offline system parameter estimation, which does not require any signal injection or additional sensor and measuring equipment, resulting in a noninvasive approach. A DT model that includes the inverter-driven PMSM power stage and a closed-loop controller is established. The key parameters of the PMSM are obtained with a chaotic PSO algorithm.
(viii) Kuprat {\em al.} \cite{Kuprat2024} proposed thermal DTs, which create a real-time-capable digital replica of the physical thermal behavior and enable online monitoring of the thermal impedance. PSO and the dual extended Kalman filter (DEKF) are used to extract the thermal model for online monitoring.
(ix) Liu {\em et al.} \cite{Liu2024} proposed a one-cycle DT method to simultaneously identify multiple component parameters of a buck converter using only several sampling points of voltage and current per switching period. PSO is employed to constantly update the component parameters. The DT can simultaneously identify parameters of inductor, capacitor, switch, and load.
(x) Nazif {\em et al.} \cite{BehzadNazif2025} analyzed the effect of sensor errors on parameter identification and eventually on DT performance. Sensor uncertainties are considered as unknowns in the parameter identification algorithm. The interior point method (IPM) and PSO are considered for parameter fitting, the first being found to be superior.
(xi) Zheng {\em et al.} \cite{Zheng2025} designed a model-driven numerical solver to improve computational efficiency, while a parameter-identification-based reconfigurable model ensures robust state tracking under parameter drift. Cognitive DT-generated information (CDT) is seamlessly integrated with model predictive control (MPC) to achieve high control performance without high-bandwidth sensors. 
\begin{table*}[t]
\caption{Comparative overview of prior works versus the proposed approach. Further details are provided in the text.}
\label{comparison}
\centering
\begin{tabular}{lllllll}
\toprule
Publication & PEC topology & Parameter identification & Task & Connectivity & Wiring & Networking\\
\midrule
\cite{Milton2020} & Buck & PCE & Diagnostics monitoring & EDT & Conventional & None \\ %DONE
\cite{Peng2021} & Buck & PSO & Health indication & EDT & Conventional & None \\ %DONE
\cite{Zhao2022} & Buck & PIML & Parameter identification & EDT & Conventional & None \\ %DONE
\cite{Lei2023} & Buck & KF & Monitoring and control & EDT & Conventional & None \\ %DONE
\cite{DiNezio2023} & Boost & PSO & Health status monitoring & EDT & Conventional & None \\ %DONE
\cite{deLopezDiz2023} & 3-phase NPC & PSO + GA & Condition monitoring & EDT & Conventional & None \\ %DONE
\cite{Song2024} & 3-phase inverter & PSO & Control and monitoring & EDT & Conventional & None \\ % DONE
\cite{Kuprat2024} & Buck & PSO + DEKF & Thermal parameter identif. & EDT & Conventional & None \\ % DONE}
\cite{Liu2024} & Buck & PSO & Parameter identific. \& degrad. monitoring & EDT & Conventional & None \\ %DONE
\cite{BehzadNazif2025} & Buck & IPM and PSO & Analysis of sensor errors & EDT & Conventional & None \\ %DONE
\cite{Zheng2025} & MAB & CDT-MPC & Predictive control & EDT & Conventional & None \\ %DONE
\cite{Wang2026b} & Buck & Hybrid RLS & Condition monitoring & EDT & Conventional & None \\ %DONE
This work & Buck & KF bank & Data detection and parameter identification & RDT & TPC-based & Optional \\ %DONE
\bottomrule
\end{tabular}
\end{table*}
(xii) Wang {\em et al.} \cite{Wang2026b} proposed an embedded DT framework tailored for real-time condition monitoring of DC--DC converters. The proposed approach employs a hierarchical DT architecture, where the low-layer executes the control logic, while the upper-layer manages iterative updating of the DT model using a hybrid recursive least squares (hybrid-RLS) parameter identification algorithm in combination with PSO.
A common thread across these publications, summarized and compared with the proposed work in TABLE~\ref{comparison}, is that external communication infrastructure is assumed (whether explicitly mentioned or not), which fundamentally sets them apart from our approach.

The main objective of our work is to establish a communication link between a PEC and the corresponding DT without external communication infrastructure--a goal achieved via the emerging paradigm of Talkative Power Conversion (TPC) \cite{Popovski2016, He2020, Liserre2023, Wang2023, PIEEE2025}. 

Switched-mode PECs naturally generate electrical switching signatures, i.\,e., ripple voltages and currents, which are traditionally suppressed by filtering. 
In contrast, by intentionally modulating the switching ripple using existing degrees of freedom (such as switching frequency, pulse width, or pulse phase), TPC permits simultaneous power and information transfer \cite{Stefanutti2008, Wu2015}. 
This concept, when applied to the communication link between PEC and DT, offers several advantages: (i) the PEC-fed power line can be reused for data transmission that is exploited by the DT; (ii) no extra hardware is required for information transmission, assuming the existing microcontroller that handles power control has sufficient capacity; (iii) the PEC and DT can be operated remotely.
In the case of employing a digital modulation scheme, a bit stream is transmitted that can cover any information. 
This allows converters to exchange telemetry data such as operational status, control commands, and diagnostic data over the shared electrical infrastructure, enabling tighter integration of energy transfer and communication across applications such as power grids, renewable energy systems, battery management systems, and motor drives. 
As an energy-efficient alternative to digital modulation, analog modulation can also be used for information transmission \cite{Moench2023}. 
Further prior work on TPC related to this paper includes signal and system modeling methods \cite{Hoeher2021, Mietzner2025, Hoeher2025Shannon, Hoeher2025multiport, Hoeher2026a, Chakraborty2026, Mewis2026b}, cyber-physical aspects \cite{Mewis2026a, Yoon2026}, and machine learning approaches \cite{Mousavi2024}.
The ability to operate the PEC and DT at distinct physical locations--made possible by TPC--promises benefits such as reduced radio frequency radiation compared to PLC and wireless communication systems \cite{Hoeher2026b}, and networked coordination among multiple PECs. 

However, the joint design of power and communication functionality introduces fundamental trade-offs. 
In particular, the switching ripple must be small enough to satisfy power-quality and efficiency requirements, yet large enough for reliable bit detection \cite{PIEEE2025}. 
This trade-off is further affected by noise, interference, varying grid or load impedances, and hardware imperfections. 
Consequently, TPC requires a joint treatment of PE performance and communication reliability. 
In this context, DT technology provides a promising framework to model, predict, and optimize the coupled cyber-physical behavior of such systems. 
This yields a mutual benefit: the DT enhances the reliability of TPC-based bit transmission, whereas TPC facilitates joint information and power transfer, which is beneficial for the DT deployment. 
Essentially, the TPC data link continuously updates the DT with real-time PEC-related sensor data, even if the DT and PEC are not at the same site. 
Original contributions and corresponding research problems include: 
\begin{itemize}
\item A new system concept is being established, whose fundamental innovation centers on employing TPC as a hardware-friendly joint communication and power link between a PEC and a corresponding DT. As opposed to the classical embedded DT scenario, remote as well as networked DT scenarios are possible without external communication infrastructure.  
\item The proposed system concept is modeled by a state-space model with emphasis on a remote DT scenario. Kalman filtering, optionally combined with a multi-hypothesis method for handling competing bit hypotheses and with an augmented state vector for parameter estimation, is studied for bit recovery and parameter estimation. 
The worst-case scenario is examined, in which only the ripple voltage is measurable, while all other parameters like the inductor current are hidden. 
\item Computer simulations and lab-based experimental validations are provided for three case studies to confirm the feasibility of the proposed concept. In contrast to related work, which relies on expensive FPGA-based DT modeling--a prevalent yet costly choice in embedded systems--our solution runs entirely on a single microcontroller. Still, processing is done on a microsecond timescale. Simulations and experimental results match well. The numerical results demonstrate reliable bit detection and fast parameter estimation. For benchmarking, the bit error rate is compared with optimum coherent detection and with threshold detection, respectively. 
\end{itemize} 

The remainder is organized as follows. 
In Section~\ref{sec:concept}, the proposed system concept is presented. 
Time-domain system modeling considering the PEC and the DT is performed in Section~\ref{sec:modeling}. 
Simulation results and experimental results are reported in Section~\ref{sec:experimental}. 
Finally, conclusions are drawn in Section~\ref{sec:conclusion}. 

\section{System Concept}\label{sec:concept} 
Recall that a DT is a virtual replica of a physical system that is continuously updated with real-time data. 
Subsequently, the physical system is assumed to be a PEC together with its control unit. 
In this context, the virtual replica is used to simulate, predict, and improve the performance of its physical counterpart. 

\begin{figure}[t]
\centering
\begin{subfigure}{0.28\textwidth}
\includegraphics[width=\linewidth]{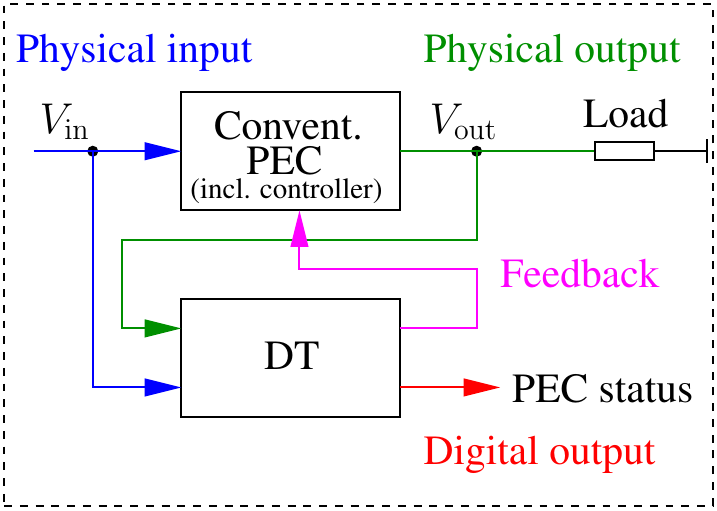}
\caption{Embedded DT}
\end{subfigure}

\vspace{3mm}

\begin{subfigure}{0.28\textwidth}
\includegraphics[width=\linewidth]{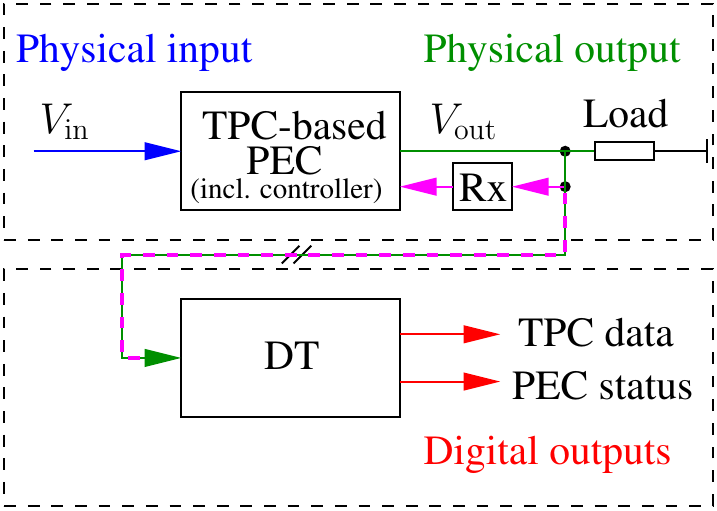}
\caption{Remote DT}
\end{subfigure}

\vspace{3mm}

\begin{subfigure}{0.28\textwidth}
\includegraphics[width=\linewidth]{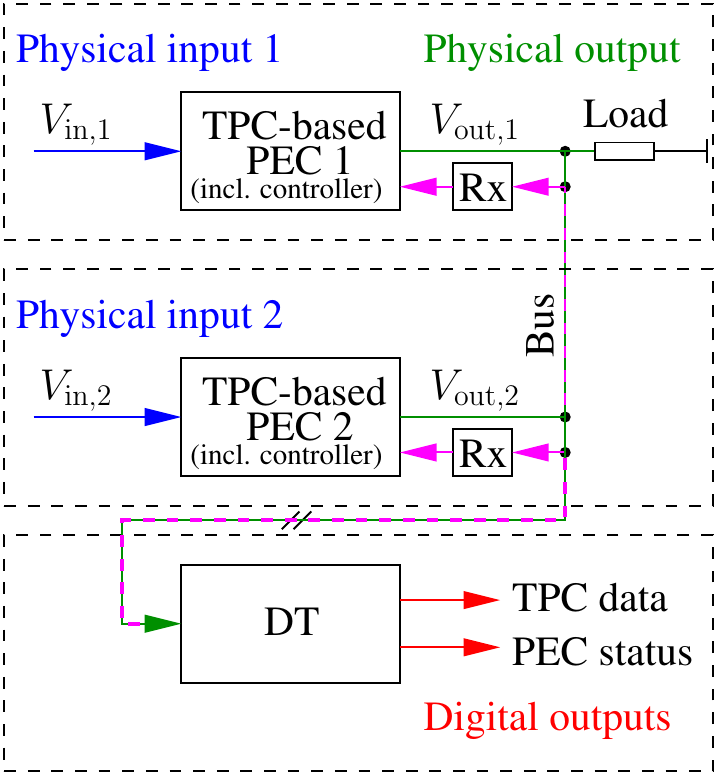}
\caption{Networked DT}
\end{subfigure}
\caption{System concepts: (a) embedded DT, (b) remote DT, (c)~networked DT. ``TPC data'' refers to the bits that are transmitted via the TPC forward link and detected in the DT. ``PEC status'' refers to PEC status data that is computed in the DT based on the TPC data.}
\label{fig:dt}
\end{figure}

The state-of-the-art and two variations of the proposed system concept are depicted in Fig.~\ref{fig:dt}. 
Conventionally, a single physical asset and the DT are co-located as shown in Fig.~\ref{fig:dt}\,(a). 
Wiring enables fast closed-loop interaction. 
This setup is referred to here as {\em embedded digital twin} (EDT) scenario. 
In Fig.~\ref{fig:dt}\,(b), a setup called {\em remote digital twin} (RDT) is depicted. 
The use of TPC eliminates the need for additional wiring.  
In the third setup illustrated in Fig.~\ref{fig:dt}\,(c), multiple physical devices are controlled by a centralized DT. 
This setup is an instance of a {\em networked digital twin} (NDT) scenario.  
In all cases, the DT provides simulated output data (also referred to as twin predictions), such as temperature, fault detection, and aging, and optionally additional information about the network in the NDT scenario. 

In the EDT scenario, the DT is connected both to the physical input and output (and, if applicable, has access to additional sensor data), while the physical device is connected to the DT via a feedback channel, see Fig.~\ref{fig:dt}\,(a). 
Typically, the feedback channel feeds the controller rather than the PEC itself \cite{Milton2020}. 
The data samples exchanged between the physical output and the DT, and between the DT and the PEC, may be quantized, but additional noise can be ignored when both units are co-located. 
All prior works reported in the introduction consider the EDT scenario. 

When the DT and PEC are co-located, information exchange via dedicated communication buses is easier and faster than through TPC, making TPC often unnecessarily complicated in the EDT scenario. 
TPC-based PEC may be only advantageous in distributed or communication-constrained systems.

In the RDT scenario, neither the DT is connected to the physical input nor the PEC is connected to the DT via a separate feedback channel, see Fig.~\ref{fig:dt}\,(b). 
To ensure the same DT functionality as in the EDT scenario, we propose to exploit the TPC-based data communication capability, provided that additional communication hardware and wiring is to be avoided.
However, possible distortions on the power line and its length affect both data transmission and parameter estimation. 
Hence, besides quantization noise, measurement noise may affect the samples exchanged between the physical output and the DT. 
For TPC the latency is small--though not negligible--and the bit rate is limited. 

Regarding the RDT scenario, two options are suggested: 
(i) A shadow mode with one-way coupling is obtained for unidirectional TPC, i.\,e., only a TPC-based forward link is established. 
The information sequence can be represented by the data-modulated output ripple voltage of the PEC. 
(ii) For two-way coupling bidirectional TPC is necessary.  
Both directions--PEC to DT (forward) and DT to PEC (feedback)--are carried over the same power lines. 
This is an all‑in‑one solution:
In the forward link, the PEC embeds its sensor data (voltage, current, temperature, etc.) into the power waveform.
In the feedback link, the DT embeds control commands into the power waveform, e.\,g.~by load modulation or by a low-cost ripple injector. 
The ripple injector could be, for example, an AC-coupled half-bridge that is phase/frequency/pulse-width modulated with feedback data. 
In this option, the PEC must be equipped with a receiver that is able to decode the bit stream sent by the DT. 
The rationale of TPC is to eliminate separate channels entirely. 
Existing power wires can be used for both power delivery and the entire data transfer.

The NDT scenario is an extension of the RDT scenario considering multiple physical assets (here: PECs) that communicate over a network (here: a power grid). 
This includes special cases like a centralized DT, a distributed DT, and hybrid versions thereof. 
A centralized DT is a deployment strategy that applies a single DT connected to multiple physical assets. 
This strategy is illustrated in Fig.~\ref{fig:dt}\,(c). 
Vice versa, a distributed DT is a strategy that allocates DT functionalities across multiple nodes \cite{Rauf2025}. 
Optionally, NDTs can estimate or reconstruct the system-wide state in addition to the individual states.
TPC is a key enabler for realizing NDT scenarios. 
Crosstalk between upstream and downstream can be avoided by a duplexing scheme (such as time-division duplexing (TDD)) and between multiple PECs by a multiple-access scheme (such as time-division multiple access (TDMA)). 
For this reason, focusing on the RDT scenario is sufficient for the scope of this paper. 

The key to a successful DT is a constant, reliable transmission of high-quality data between the physical asset and the DT. 
This is where TPC provides added value:
\begin{itemize}
\item Seamless data acquisition: Traditional DTs often rely on separate sensors and communication networks to transfer this data. TPC elegantly solves this by turning the PEC into a data source. The same hardware that manages power transfer also generates the bit stream, effectively making the converter a self-reporting device.
\item Cost-effective digitalization: By eliminating the need for extra communication hardware, TPC can reduce the cost and complexity of implementing digital twins, making them more practical for a wider range of systems, from single PECs to networked PECs and from microgrids to smart grids.
\item Enabling a true cyber-physical system: The combination of a TPC-enabled PEC and a DT creates a deeply integrated system. The physical converter reports its state via "power talk", and the DT can, optionally, send control signals back. This provides a closed loop, creating a responsive and intelligent cyber-physical system.
\item Improved bit detection: As the DT establishes a precise model of the physical asset, more reliable detection of the TPC-based bit stream is possible. This is particularly important in harsh environments, for instance industrial environments, for long communication distances, and in networks that employ several PECs. Improved bit detection provides added value, even if traditional DT functionalities such as fault detection and condition monitoring are turned off or are unwanted. 
\item Integrated power transfer: The DT can be powered by the TPC-based PEC, particularly if it is remote.
\end{itemize}

In summary, the DT is not merely a KF-based observer; it is a holistic DT that facilitates continuous synchronization and rich sensor data exchange, and can potentially serve as a virtual replica beyond state estimation. 
The key enabler is TPC-based data transmission on the forward link from the PEC to the DT (especially for PEC-related sensor data), with an optional similar link in the reverse direction (primarily for PEC control data). 

In the subsequent sections on system modeling (Section~\ref{sec:modeling}) and computer simulations plus experimental validation (Section~\ref{sec:experimental}), three use cases are exemplarily studied: bit detection, adaptation to output voltage variations, and parameter estimation with emphasis on load variations. 

\section{System Modeling}\label{sec:modeling}
This section deals with time-domain system modeling. 
We begin with a state-space model assuming a synchronous buck converter as an example, before moving on to KF-based modeling of the corresponding DT. 
For real-time modeling applications such as HIL simulations or DTs, highly computationally efficient large-signal models are important that can be directly implemented in any
computational environment, including directly on an existing embedded controller \cite{Wunderlich2023}. 

\begin{figure}[t]
\centering
\includegraphics[width=0.8\linewidth]{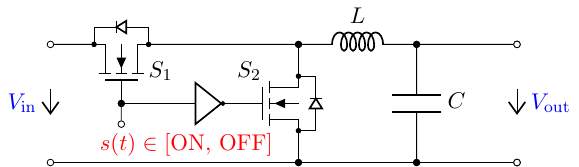}
\caption{Topology of the synchronous buck converter used in the numerical results. The switching signal is denoted as $s(t)$.}
\label{buck}
\end{figure}

\subsection{PEC Modeling}
\begin{figure}[t]
\centering
\includegraphics[width=\linewidth]{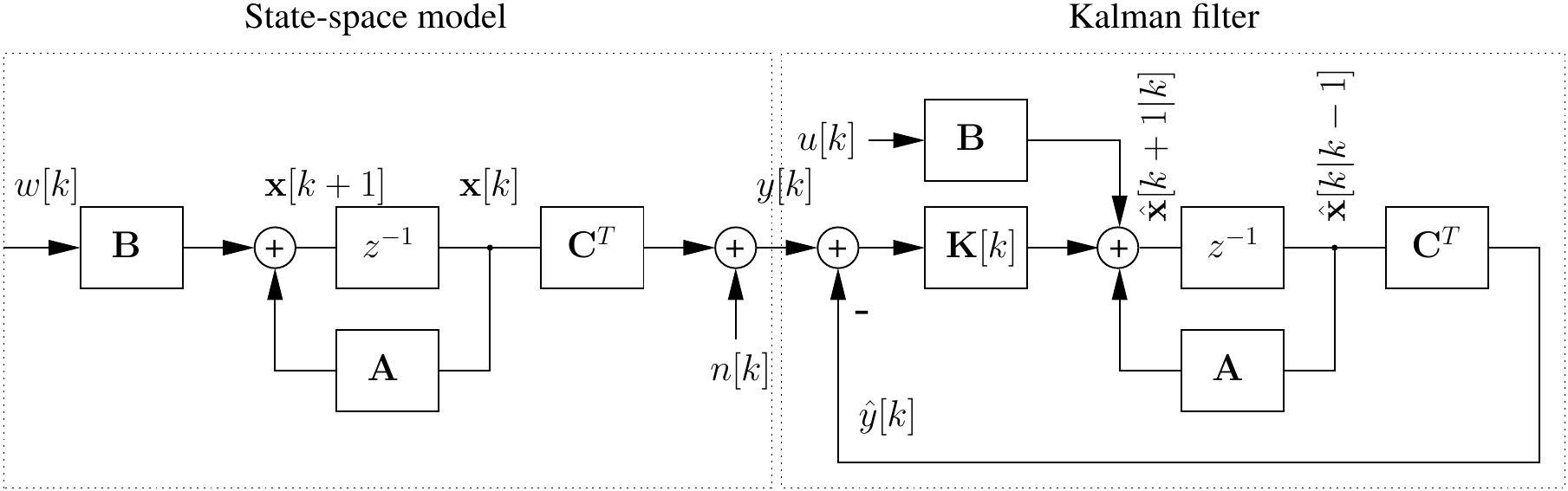}
\caption{Block diagram of the considered state-space model and the corresponding Kalman filter.} 
\label{fig:kalman}
\end{figure}

The numerical model of the PEC described below is based on the fact that ideal inductors $L$ and capacitors $C$ can be represented by differential equations: 
\begin{align}
L\, \frac{di_\mathrm{L}(t)}{dt} &= v_\mathrm{L}(t)\\
C\, \frac{dv_\mathrm{C}(t)}{dt} &= i_\mathrm{C}(t).
\end{align}
For the example of a synchronous buck converter as depicted in Fig.~\ref{buck}, these differential equations can be approximated by difference equations as \cite{Hoeher2021, Mietzner2025}
\begin{align}
L\, \frac{i_\mathrm{L}[k+1]-i_\mathrm{L}[k]}{\Delta T} &= \underbrace{s[k] v_\mathrm{in}[k] - v_\mathrm{out}[k]}_{v_\mathrm{L}[k]}\label{1}\\
C\, \frac{v_\mathrm{out}[k+1]-v_\mathrm{out}[k]}{\Delta T} &= \underbrace{i_\mathrm{L}[k] - \frac{v_\mathrm{out}[k]}{R}}_{i_\mathrm{C}[k]},\label{2}
\end{align}
where $k$ is the time index, $\Delta T$ is the time step associated with forward Euler approximation, $s[k]\in\{0,1\}$ is the sampled switching signal (1 = high-side closed, 0 = low-side closed), $v_\mathrm{in}[k]$ is the input voltage of the PEC, $i_\mathrm{L}[k]$ is the inductor current, $v_\mathrm{out}[k]=v_\mathrm{C}[k]$ is the output voltage of the PEC, and $R$ is the load resistance.
Convergence between analog and discretized modeling is achieved if the time step $\Delta T$ is much shorter than the switching period \cite{Mietzner2025}.  
The switching signal $s[k]$ is modulated by pulse-width modulation (PWM). 
In the case of TPC, $s[k]$ is bit-dependent. 
We assume that $K_\mathrm{Tx}$ samples are taken per switching period $T_\mathrm{sw}$, e.\,g., $K_\mathrm{Tx}=1024$. 
The relation between these parameters is $\Delta T = T_\mathrm{sw}/K_\mathrm{Tx}$. 
The bit rate is maximized, if the symbol duration $T$ is the same as $T_\mathrm{sw}$. 
This case is assumed subsequently. 
Given a constant duty cycle $0<D<1$, the relation between the average input voltage $V_\mathrm{in}=E\{v_\mathrm{in}[k]\}$ and the average output voltage $V_\mathrm{out}=E\{v_\mathrm{out}[k]\}$ of the DC--DC step-down PEC is $V_\mathrm{out}=D\cdot V_\mathrm{in}$, where $E\{.\}$ denotes the expected value operator. 
The time-varying output voltage can be decomposed as $v_\mathrm{out}[k]=V_\mathrm{out}+v_\mathrm{rip}[k]$. 
Rearranging (\ref{1}) and (\ref{2}) yields the state-space model illustrated on the left-hand side of Fig.~\ref{fig:kalman}: 
\begin{align}
{\bf x}[k+1] &= {\bf A}\, {\bf x}[k] + {\bf B}\,w[k] \label{x} \\
{\bf y}[k] &= {\bf C}^T\, {\bf x}[k] + {\bf n}[k] \label{y}
\end{align}
with state vector
\begin{equation}
{\bf x}[k] = \left[i_\mathrm{L}[k], v_\mathrm{out}[k]\right]^T,
\end{equation}
PWM-signal-dependent input signal of the LC filter 
\begin{equation}\label{w}
w[k] = s[k]\cdot v_\mathrm{in}[k],
\end{equation}
system matrix 
\begin{equation}\label{A}
{\bf A} = 
\begin{bmatrix}
1 & -\Delta T / L\\
\Delta T / C &  1-\Delta T / (RC)
\end{bmatrix},
\end{equation}
input matrix 
\begin{equation}\label{B}
{\bf B} = \left[\Delta T / L, ~0\right]^T,
\end{equation}
and measurement matrix
\begin{equation}\label{C}
{\bf C} = {\bf I}_2,
\end{equation}
where ${\bf I}_2$ is the $2\times 2$ identity matrix. 
In the EDT scenario, $s[k]$ is known and $i_\mathrm{L}[k]$, $v_\mathrm{in}[k]$, and $v_\mathrm{out}[k]$ are measurable. 
The elements of the noise vector ${\bf n}[k]$ are typically small. 

If the inductor current $i_\mathrm{L}[k]$ is a hidden variable (i.\,e., not directly measurable) and $s[k]$ is unknown, like in the RDT scenario, the discrete-time state-space model remains the same in terms of dynamics, but the measurement equation only includes the measurable output voltage. 
Then, ${\bf C}=[0, 1]^T$ and consequently 
\begin{equation}
y[k] = v_\mathrm{out}[k]+n[k], \label{yk}
\end{equation}
where the noise variance $\sigma_n^2=E\{n^2[k]\}$ may not be negligible any more in the case of remote measurements. 
According to the state matrix ${\bf A}$, the coupling between $i_\mathrm{L}[k]$ and $v_\mathrm{out}[k]$ is weak. 
%The coupling can be increased by raising $\Delta T$.

In the RDT and related scenarios, even for random TPC data it may happen that the input signal $w[k]$ defined in (\ref{w}) is partly known. 
As can be seen in Fig.~\ref{fig:fsk}, some chips of the switching signals $s_0(t)$ (for bit=0) and $s_1(t)$ (for bit=1) are identical for a given duty cycle. 
This situation can be modeled as 
\begin{equation}\label{wk}
w[k] = u[k] + v[k] = s[k]\, v_\mathrm{in}[k]
\end{equation}
with
\begin{equation}\label{vk}
u[k] = \overline{s}[k]\, v_\mathrm{in}[k] \quad \mathrm{and} \quad  
v[k] = \tilde{s}[k]\, v_\mathrm{in}[k],
\end{equation}
where $\overline{s}[k]\in \{0, 0.5, 1\}$ is the deterministic part of the switching signal subject to $E\{\overline{s}[k]\}=D$, while $\tilde{s}[k]\in \{-0.5, 0, 0.5\}$ is the unknown part of the switching signal. 
Consequently, it can be assumed that the deterministic input signal $u[k]$ is known at the receiver. 
In contrast, the zero-mean input signal $v[k]$ is unknown and must be estimated. 
Its variance is given by $\sigma_v^2[k] = E\{v^2[k]\} = v_\mathrm{in}^2[k]/4$ for time indices $k$ where $\overline{s}[k] = 0.5$, and $\sigma_v^2[k] = 0$ otherwise.
Equations (\ref{yk})--(\ref{vk}) extend the state-space modeling published for example in \cite{Milton2020, Lei2023}.

The system model of the buck converter presented so far is affected by the power control strategy via (\ref{B}). 
The instantaneous duty cycle can be expressed as $d[k]=\mathrm{avg}\{s[k]\}$, where $\mathrm{avg}\{.\}$ denotes the short-term averaging operator, subject to $D=E\{d[k]\}$.  

\subsection{DT Modeling}
A key task of the DT is parameter estimation to reproduce the behavior of the PEC.
To this end, a state-space model of the PEC under investigation was derived above.
Due to parameter uncertainties and the possibility of load changes, the DT must be adaptive.
The DT can be adjusted using linear or nonlinear numerical optimization algorithms \cite{Sarkka2023}.
As outlined in the introduction, numerous identification algorithms have been investigated in the state of the art; see, for example, \cite{Milton2020}–\cite{Wang2026b}.
Given a precise state-space model, parameter identification using Kalman filtering is an obvious solution, since a KF is the optimal linear LMMSE (Linear Minimum Mean-Square Error) estimator of the state vector ${\bf x}[k]$, based on the measurements available up to time index $k$.
A standard Kalman filter matched to the given discrete-time state-space model is shown on the right-hand side of Fig.~\ref{fig:kalman}.
We have used this technology as the baseline technology in DT.
For better readability of the main text, the corresponding KF equations are collected in the appendix \cite{Anderson1979}.

For TPC-based bit transmission, a modification of the standard KF is recommended. 
However, the resulting computational complexity should remain manageable. 
Therefore, we distinguish between different cases matched to the specific task at hand and adapt the algorithms accordingly. 
In the following, three case studies are defined and evaluated for the RDT scenario. 

\subsection{Case Study I: Bit Detection}
Being a precise, adaptable copy of the physical asset, the DT is able to serve as an advanced demodulator. 
Generally speaking, TPC-based bit detection is possible by many strategies in the time domain, frequency domain, or statistical domain \cite{PIEEE2025}. 
Methods commonly applied in the time domain are threshold detection, slope detection, zero crossing, and maximum-likelihood detection. 
In the frequency domain, discrete or fast Fourier transform (DFT or FFT), Goertzel algorithm, and bandpass filtering are possible solutions. 
Machine-learning-based reception is an example of demodulation in the statistical domain. 
% Equalization: \cite{Chakraborty2026}. 
The DT outputs the bit decisions optionally together with simulated data such as temperature and power efficiency. 
Let us consider several approaches regarding bit detection.

\begin{figure}[t]
\centering
\includegraphics[width=0.8\linewidth]{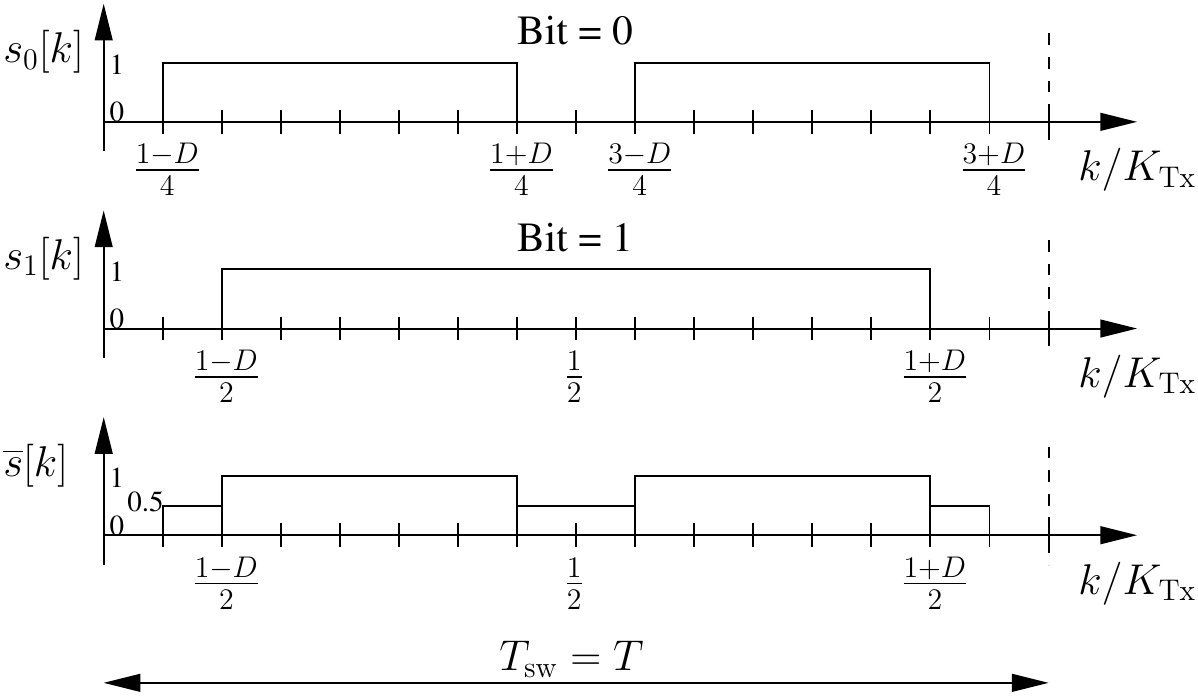}
\caption{Binary PWM-FSK with duty cycle $D=0.75$. The top and middle subfigures present the two possible hypotheses for bit=0 and bit=1, respectively, while the bottom subfigure shows the average switching signal $\overline{s}[k]$, which may serve as an input signal of a Kalman filter. The impact of this input on bit decisions is ambiguous only for time indices $k$ where $\overline{s}[k]=0.5$.} 
\label{fig:fsk}
\end{figure}
 
Option~1: A simple (but not elegant) solution is to use a conventional bit detection algorithm--operating in the time, frequency, or statistical domain--in parallel to the KF observer. 
The conventional detector could be applied to the received samples or to filtered/predicted output voltage or inductor current samples. 
In the numerical results section, Option~1 serves as the baseline for performance comparison. 

Option~2: Because only $v_\mathrm{out}[k]$ is measurable in the RDT and related scenarios, $i_\mathrm{L}[k]$ is hidden while $s[k]$ is partly unknown (except for training sequences). 
The standard two-state KF is able to estimate both the inductor current and the bit stream.
One could feed the standard KF with the average input signal $u[k] = \overline{s}[k]\, v_\mathrm{in}[k]$, see the upper left corner of the KF in Fig.~\ref{fig:kalman} and the bottom subfigure in Fig.~\ref{fig:fsk} for the case of binary PWM-FSK. 
This is preferable to omitting the KF input and increasing $\sigma_v^2=E\{v^2[k]\}$.
Still, convergence is an issue for high data rates. 

Option~3: Another possibility to estimate the partly unknown two-level input signal $w[k] = s[k]\, v_\mathrm{in}[k]$ using a standard two-state KF is to directly obtain $\hat{w}[k]$ from the state estimates. 
To this end, the estimated states are exploited at two consecutive time instants. 
The system dynamics in (\ref{x}) can be written as 
\begin{equation}
i_\mathrm{L}[k+1] = i_\mathrm{L}[k] - (\Delta T / L) \, v_\mathrm{out}[k] + (\Delta T / L) \, w[k].
\end{equation}
Solving for $w[k]$ yields
\begin{equation}
w[k] = (L/\Delta T) (i_\mathrm{L}[k+1] - i_\mathrm{L}[k]) + v_{\text{out}}[k].
\end{equation}
By replacing the true states with the filtered estimates (after measurement update) at times $k$ and $k+1$, finally
\begin{equation}
\hat{w}[k] = (L/\Delta T) (\hat{i}_\mathrm{L}[k+1|k+1] - \hat{i}_L[k|k]) + \hat{v}_{\text{out}}[k|k]
\end{equation}
is obtained. 
In this solution, the average switching signal $\overline{s}[k]$ should be taken into account, and the process noise variance $\sigma_v^2[k]$ should be large enough to account for the mismatch. 
Since the estimate $\hat w[k]$ is real-valued, a two-level value can be obtained by threshold decision. 

Notice, however, that estimates of $v[k]$ can be obtained if and only if the system is input-observable. 
Since ${\bf C}^T\,{\bf B}=0$, the input does not directly affect the output. 
This means that $v[k]$ (and hence $w[k]$) cannot be detected instantly. 
But the system is weakly observable through dynamics, because ${\bf C}^T\,{\bf A}\,{\bf B}\not=0$.
Nevertheless, since matrix element $a_{21}$ in ${\bf A}$ is small, the effect of $w[k]$ on $y[k]$ is small as well. 
Detection relies on accumulation over many samples. 
Therefore, we do not recommend this option for the purpose of bit detection. 

Option~4: If higher accuracy is targeted, more powerful approaches include \cite{Sarkka2023}:
(i) a multi-model Kalman filter bank, each using a different model, together with a hypothesis-testing algorithm; 
(ii) an augmented Kalman filter (AKF) for estimating system states and unknown parameters (for instance $s[k]$ besides $i_\mathrm{L}[k]$ and $v_\mathrm{out}[k]$); 
(iii) an extended Kalman filter (EKF) or unscented KF (UKF) for dealing with nonlinear systems--a situation that can occur if new states enter the dynamics or measurement equations in a nonlinear way;   
(iv) hybrid estimation performing KF for state estimation along with a detection strategy for $s[k]$; 
(v) maximum a-posteriori probability (MAP) estimation; 
(vi) particle filtering, etc.

\begin{figure}[t]
\centering
\includegraphics[width=0.8\linewidth]{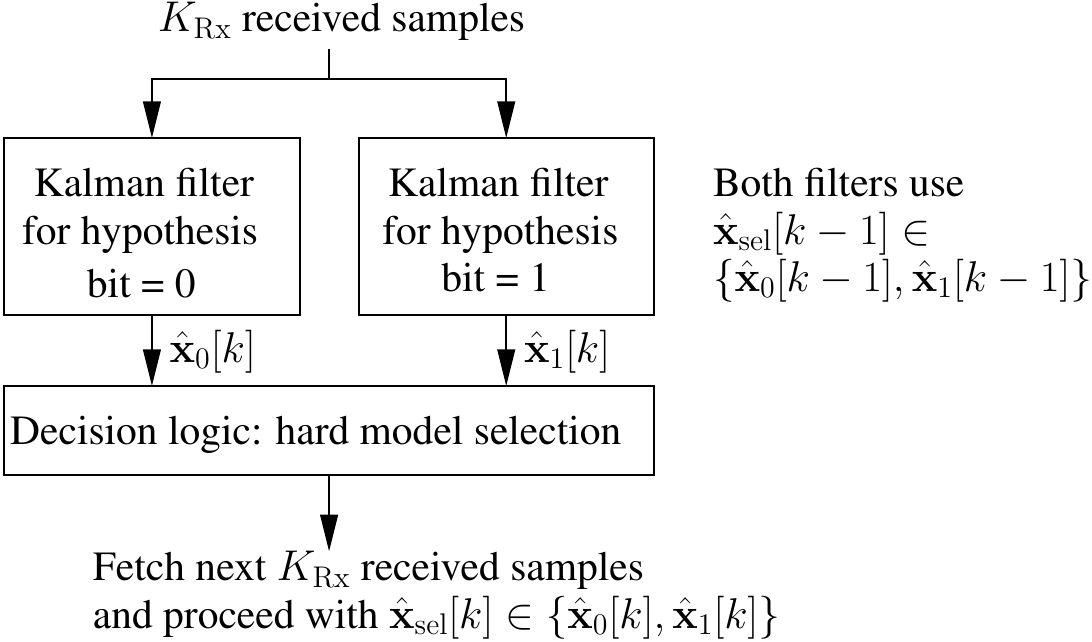}
\caption{Sketch of a two-model Kalman filter bank with hard model selection.} 
\label{fig:selection}
\end{figure}

After careful consideration and initial testing, we decided to implement a two-model KF bank \cite{Hanlon2000}. 
This solution works well for discrete-valued input symbols of small cardinality. 
Applied to the given system and ignoring intersymbol-interference (ISI) caused by LC filtering, for binary modulation schemes the KF bank method uses two models with hypotheses $s[k]=s_0[k]$ (for bit=0) and $s[k]=s_1[k]$ (for bit=1), runs two KFs in parallel, and performs hypothesis testing. 
Although not pursued further here, a multi-model KF bank could account for ISI by processing bit tuples instead of individual bits, at the cost of increased computational complexity. 
A multi-model KF bank could also be used to estimate the duty cycle. 

Regarding hypothesis testing, a variety of solutions have been published that select or weight the filter bank estimates, such as the interacting multiple-mode (IMM) method \cite{Dingler2022}. 
In the numerical examples, the simplest approach--namely, hard model selection as illustrated in Fig.~\ref{fig:selection}--was used, which does not involve the implementation of model probabilities or proper mixing.  
Low computational complexity further benefits from the fact that the error covariance matrix (and therefore also the Kalman gain) is independent of the bit hypothesis. 
Furthermore, the error covariance matrix is scalar. 
In this special case and if no a-priori information about the bits is available, hard model selection can be obtained from the sum of squared innovations as 
\begin{equation}\label{hat_w}
\hat w = \arg \min_{i\in\{0,1\}} \underbrace{\sum_{k=0}^{K_\mathrm{Rx}-1} e_i^2[k]}_{\mathrm{sum}_i},
\end{equation}
where $e_0[k]=y[k]-\hat y_0[k]$ and $e_1[k]=y[k]-\hat y_1[k]$ are the innovations for bit=0 and bit=1, respectively, and $K_\mathrm{Rx}$ (with $K_\mathrm{Rx}\leq K_\mathrm{Tx}$, e.\,g.~$K_\mathrm{Rx}=16$) is the number of receiver-side samples per switching period $T_\mathrm{sw}$, i.\,e., per symbol duration $T$ and hence per bit interval. 
Consequently, the time step $\Delta T$ is enlarged at the receiver side, which is beneficial regarding the off-diagonal elements in ${\bf A}$. 
Equation (\ref{hat_w}) is to be evaluated for each bit. 
At the end of each bit interval consisting of $K_\mathrm{Rx}$ samples, the state estimate that corresponds to the bit decision is selected and taken as a starting point for the next bit interval, see Fig.~\ref{fig:selection}. 
This solution provides precise estimates, but is simpler than likelihood-based weighting and is therefore used in the numerical results reported in Section~\ref{sec:experimental}. 

Naturally, hard model selection provides hard bit decisions. 
Interestingly, a corresponding soft bit decision can be obtained without extra effort as 
\begin{equation}\label{soft}
\omega = \mathrm{sum}_0 - \mathrm{sum}_1 .
\end{equation}
The soft bit decision $\omega$ includes the hard bit decision (\ref{hat_w}) according to $\hat w = \mathrm{sign}(\omega)$, as well as additional reliability information: the larger $|\omega|$ is, the more reliable the bit decision is. 

\subsection{Case Study II: Adaptation to Output Voltage Variations}
Output voltage variations can have several causes.  
On the one hand, they may be operational, for example, to supply different loads or to adapt the voltage during battery charging. 
On the other hand, they can be used for testing purposes to evaluate the tracking performance of the power control loop. 
For these reasons, in this case study, in the physical system the DC output voltage is periodically toggled between 24~V and 16~V by abruptly changing the duty cycle between $D_1=0.75$ and $D_2=0.5$, given a constant input voltage of 32~V. 
Each interval lasts for $2\times 16$~bit durations, i.\,e., $640~\mu$s. 
In terms of the DT algorithms, no changes were made compared to Case Study~I.

\subsection{Case Study III: Parameter Estimation of PE Components} 
So far, the values of all PE components are assumed to be known in the DT and they are assumed to be ideal. 
In practice, however, these components may be time-varying, not known precisely, and/or subject to imperfections. 
Reasons for time-variations are aging and changes of the point of operation, like load changes, inductor saturation, and the DC-bias dependence of capacitors. 
Particularly the load $R$ is often subject to noticeable variations. 
Therefore, an investigation of load changes is targeted in this final case study. 
In this case study, in the physical system the load is toggled between $R_1=3.3~\Omega$ and $R_2=2.5~\Omega$ by periodically adding/removing a $10~\Omega$ resistor in parallel to the nominal load resistor, given a constant PEC input voltage of 32~V.
Each interval lasts for $2\times 16$~bit durations, i.\,e., $640~\mu$s. 

Due to the possibility of abrupt load changes, we opted to implement a two-model AKF filter bank in combination with hard model selection \cite{Hanlon2000}. 
The AKF augments the state vector with at least one additional variable \cite{Anderson1979}.
Here, the state vector is augmented by the time-varying load conductance $G[k]=1/R[k]$, i.\,e.,
\begin{equation} 
{\bf x}[k]=[i_\mathrm{L}[k], v_\mathrm{out}[k], G[k]]^T. 
\end{equation}
The load conductance is modeled as a random walk:
\begin{equation}
G[k+1] = G[k] + q_g[k], 
\end{equation}
where $q_g[k]$ is a zero-mean process noise sample. 
The resulting state-transition model is nonlinear because the capacitor equation contains the product $G[k]\,v_\mathrm{out}[k]$. 
Therefore, an extended Kalman filter (EKF) is employed. 
The nonlinear state-transition function is given by 
\begin{equation} 
{\bf x}[k+1] = {\bf f}({\bf x}[k], w[k]) + {\bf q}[k], 
\end{equation} 
where 
\begin{equation} 
{\bf f}({\bf x}[k], w[k]) = 
\begin{bmatrix} i_\mathrm{L}[k]+\dfrac{\Delta T}{L} \left(w[k]-v_\mathrm{out}[k]\right) \\ v_\mathrm{out}[k]+\dfrac{\Delta T}{C} \left(i_\mathrm{L}[k]-G[k] v_\mathrm{out}[k]\right) \\ G[k] 
\end{bmatrix}. 
\end{equation} 
The Jacobian of the state-transition function, evaluated at the current state estimate $\hat {\bf x}[k|k]$, is 
\begin{align} 
{\bf F}[k] &= \left. \frac{\partial{\bf f}}{\partial{\bf x}} \right|_{\hat{\bf x}[k|k]}\nonumber \\ 
&= 
\begin{bmatrix} 1 & -\frac{\Delta T}{L} & 0\\ \frac{\Delta T}{C} & 1-\frac{\Delta T}{C} \hat G[k|k] & -\frac{\Delta T}{C}\hat v_\mathrm{out}[k|k]\\ 0 & 0 & 1 
\end{bmatrix}. 
\end{align} 
The input and measurement matrices are 
\begin{equation} 
{\bf B} = \left[\Delta T/L, 0, 0\right]^T, \qquad {\bf C} = [0, 1, 0]^T. 
\end{equation} 
The EKF prediction and update equations follow the standard Kalman-filter formulation, with the nonlinear state-transition function used for state prediction and the Jacobian $F[k]$ used for covariance prediction. 
The process-noise covariance matrix ${\bf Q}$ is of dimension $3\times 3$. 
An example is reported in the numerical results section. 

The AKF can also be used to estimate other unknown or time-varying PE components that are subject to tolerances, nonlinear effects, or aging. 
Research regarding the influence of hardware imperfections on the modeling of TPC systems has recently been initiated in \cite{Mietzner2025}.

\section{Computer Simulations and Experimental Validation}\label{sec:experimental}
\begin{figure}[t]
\centering
\includegraphics[width=\linewidth]{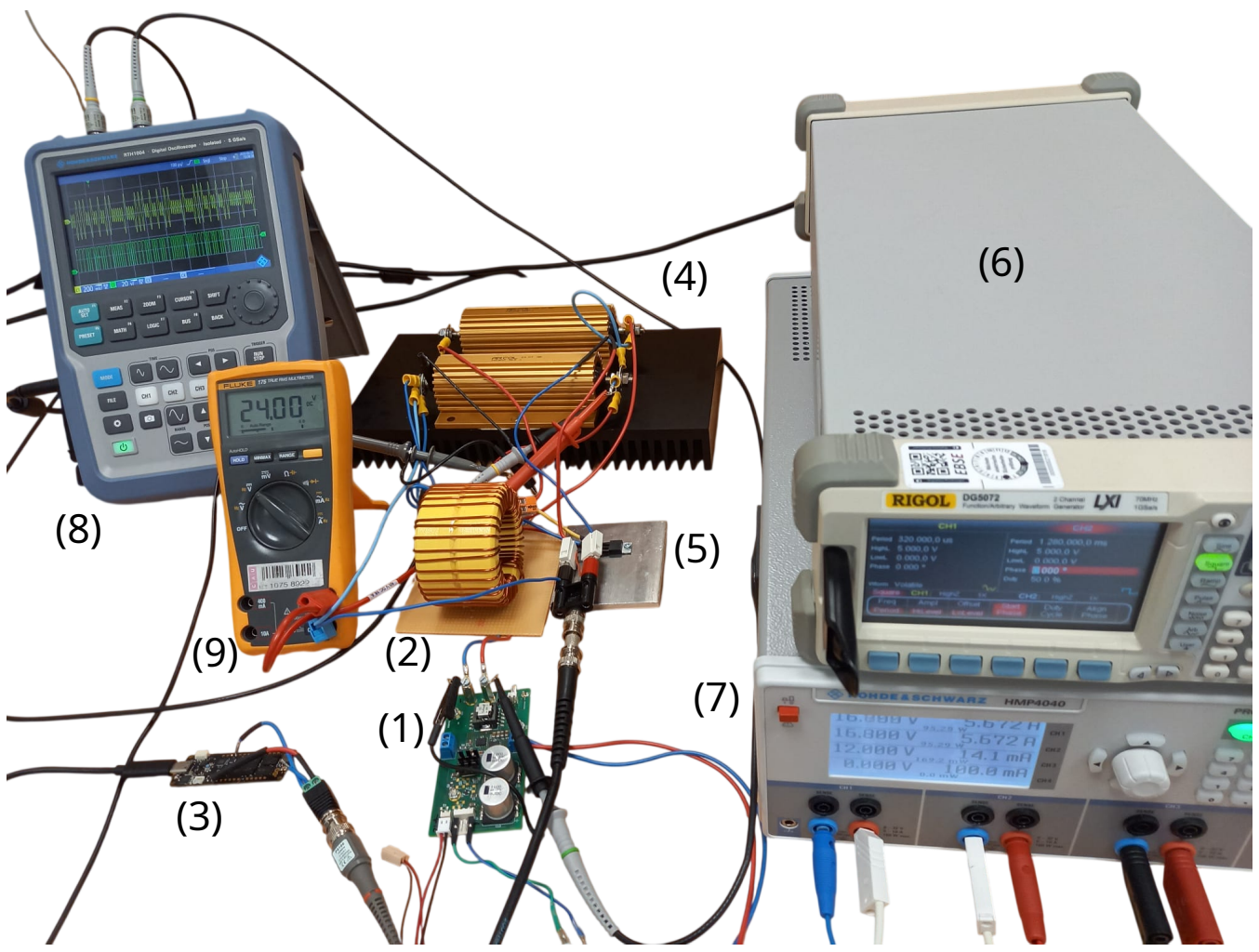}
\caption{Photo of the experimental setup. (1) Half-bridge, (2) LC filter, (3) STM32~H7 microcontroller, (4) load, (5) load switch, (6) arbitrary waveform generator, (7) power supply, (8) oscilloscope, (9) multimeter. In the ``remote DT'' scenario under investigation, the Kalman filter online computations were performed in the microcontroller. The coaxial cable connecting the STM32 board's ADC to the load can be seen at the bottom right of the STM32 board--this is the remote bus. A notebook used for simulations and offline computations is not shown here. It is connected via the USB cable on the left-hand side of the STM32 board.}
\label{experiment}
\end{figure}

Based on the system design proposed in the previous sections, this section provides numerical results obtained by simulations and by lab-based experiments, both for steady-state and transient behaviors. 
The experiments are used to validate the accuracy of the proposed time-domain modeling with emphasis on the high signal-to-noise ratio (SNR) region. 
Additional simulation results also explore the influence of noisy channel conditions on bit errors and parameter estimation accuracy. 
The nominal parameters that are selected for the investigations are listed in TABLE~\ref{parameters}. 
By focusing on one PEC topology, operating point, and bit rate, the results of the case studies are comparable.
A photo of the experimental setup is shown in Fig.~\ref{experiment}.

\begin{table}
\caption{Setup and nominal parameters used in computer simulations and experimental validation. In the Case Studies II and III, selected nominal parameters are subject to change.}
\label{parameters}
\centering
\begin{tabular}{ll}
\toprule
Parameter & Value \\ 
\midrule
PEC topology & Buck converter \\
Switching frequencies $f_0$, $f_1$ & 100~kHz (bit=0), 50~kHz (bit=1) \\
Inductance $L$ & 470~$\mu$H (Fastron TLC/10A) \\ % effective: 360uH
Output capacitance $C$ & 1~$\mu$F (WIMA MKS-4, 100V-) \\
Ohmic load $R_{\mathrm L}$ & 3.3~$\Omega$ (Arcol 300 3R3) \\
DC input voltage $V_\mathrm{in}$ of PEC & 32~V \\
DC output voltage $V_\mathrm{out}$ of PEC & 24~V \\
Average duty cycle $D$ & 0.75 \\
Power/data modulation scheme & PWM-FSK \\
Symbol duration $T$ & 20~$\mu$s\\
Raw bit rate & 50~kbps \\
Peak-to-peak ripple voltage w/o TPC & 0.20~V @ 24~V DC output \\
Peak-to-peak ripple voltage with TPC & 0.66~V @ 24~V DC output \\
Peak inductor current with TPC & 7.44~A \\
\bottomrule
\end{tabular}
\end{table}

In all experiments, a custom-made synchronous buck converter with MOSFET half-bridge according to Fig.~\ref{buck} was employed with disabled power control \cite{Hoeher2021}. 
Because the half-bridge circuit can switch above 1~MHz, the signal fed to the LC filter was approximately square-wave. 
The LC filter of the PEC was implemented using a $1~\mu$F polyester film capacitor, which exhibits very little capacitance change with DC bias, together with a $470~\mu$H toroidal line choke with an iron dust core. 
Its inductance at the operating current and frequency was approximately $360~\mu$H -- a value that is assumed in the following analysis.
For the effective parameters, the damping factor of the 2nd-order LC lowpass filter is $\zeta\approx 3$, i.\,e., the filter is overdamped. 
The $10-90$\% rise/fall time of the step response is about $300~\mu$s. 
DC power supply was provided by a programmable 4-channel power source R\&S HMP4040. 
The input voltage $V_\mathrm{in}$ of the PEC was kept constant in all experiments, i.\,e., $v_\mathrm{in}[k]=V_\mathrm{in}$ for all $k$. 
The desired ratio between the DC input/output voltages $V_\mathrm{in}$ and $V_\mathrm{out}$ was adjusted by the duty cycle $D$. 
As a modulation scheme, binary 2-level FSK (also called binary PWM-FSK) was chosen due to its insensitivity to model uncertainty. 
All FSK-modulated switching signals were generated by an Analog Discovery Pro 3000 arbitrary waveform generator (AWG) based on the bit sequence $[1, 0, 0, 1, 1, 1, 0, 1, 0, 1, 0, 1, 1, 0, 0, 0]$. 
For the sake of repeatability, the bit sequence was cyclically extended. 

Computer simulations were performed in ANSI~C, while experimental results were obtained in two versions: online and offline. 
Online experiments were obtained for an Arduino Portenta microcontroller employing an STM32 H7 kernel. 
The STM32~H7 series is popular in high-performance applications due to a 32-bit Arm Cortex-M7 MCU (with some dual-core M7+M4 variants) operating at up to 600~MHz, delivering over 3300 CoreMark with a double-precision FPU, featuring up to 2~MB of Flash and 1.4~MB of SRAM, extensive connectivity, high-resolution timers, fast 16/12-bit ADCs, security accelerators, and is suitable for industrial temperature ranges--but a fast digital signal processor (DSP) or an FPGA board could also be used. 
The sampling frequency of the built-in ADC was chosen as 800~kSa/s, i.\,e., 16 samples were taken per $20~\mu$s switching period $T_\mathrm{sw}$. 
This is the minimum number of samples per symbol duration required to realize PWM-FSK modulation without distortion for $D=3/4$, see Fig.~\ref{fig:fsk}. 
The sampling theorem is fulfilled for both switching frequencies. 
The DC plus ripple output voltage has been scaled down by a 10:1 voltage divider (420~$\Omega$, 47~$\Omega$) before board-internal 14-bit analog-to-digital conversion took place. 
A 15~nF ceramic capacitor connected in parallel to the output port of the voltage divider provides lowpass filtering with a cutoff frequency of 225~kHz.
This bandwidth offers a good compromise between suppressing noise and spikes on the one hand, and preventing signal distortion and aliasing on the other. 
Regarding offline experiments, receiver-side sampling and lowpass filtering were carried out with an R\&S RTH1004 digital oscilloscope at 2.5~MSa/s, with a 200~kHz lowpass filter enabled.

Next, numerical results obtained by simulations and experiments are reported for the three case studies defined in the previous section.
In all results, the inductor current is assumed to be a hidden state variable, i.\,e., (\ref{yk}) applies for the observations. 

\begin{figure}[p]
\centering
\begin{subfigure}{0.485\textwidth}
\includegraphics[width=\linewidth]{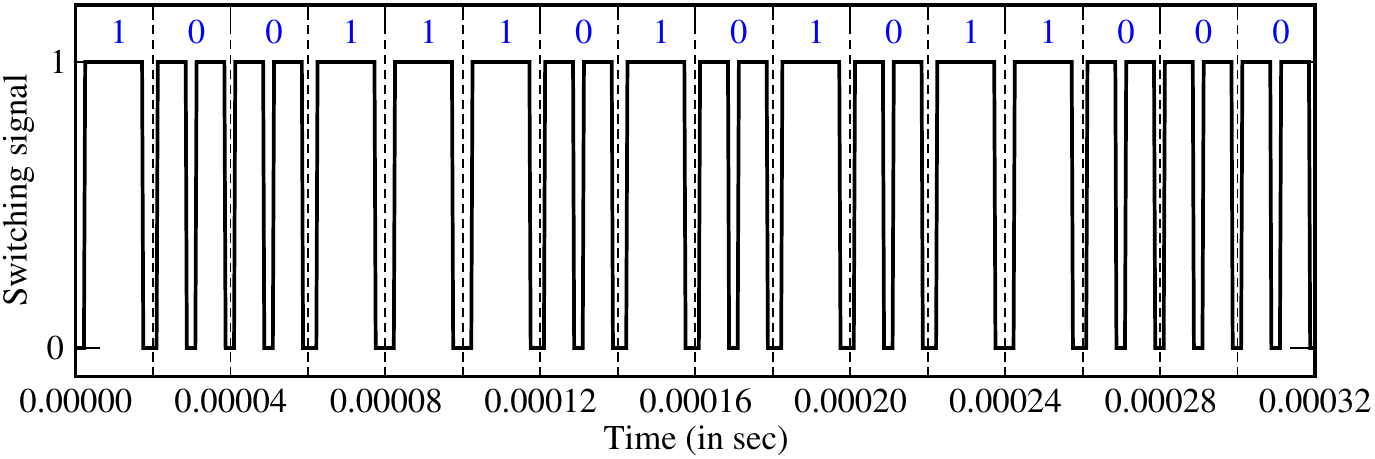}
\caption{FSK-modulated switching signal.}
\end{subfigure}

\vspace{3mm}
\begin{subfigure}{0.485\textwidth}
\includegraphics[width=\linewidth]{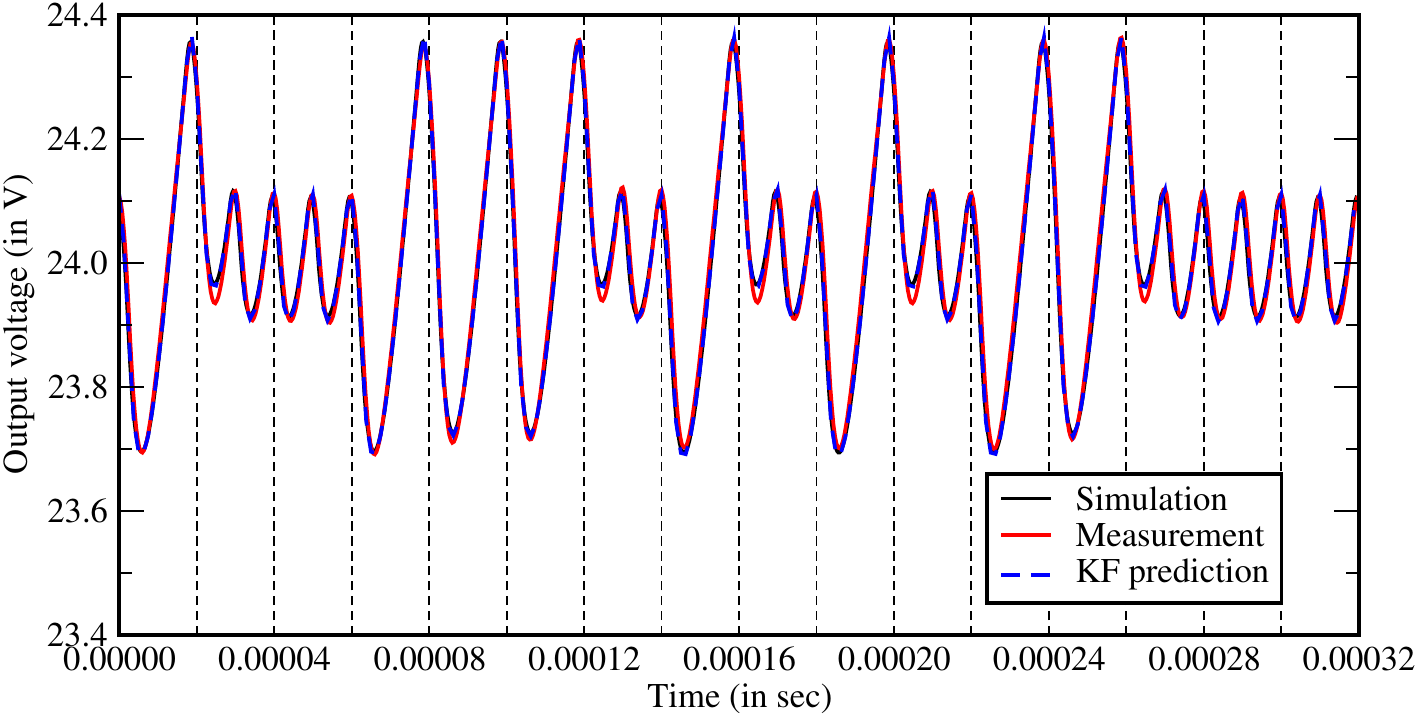}
\caption{Simulated and measured output voltage.}
\end{subfigure}

\vspace{3mm}
\begin{subfigure}{0.485\textwidth}
\includegraphics[width=\linewidth]{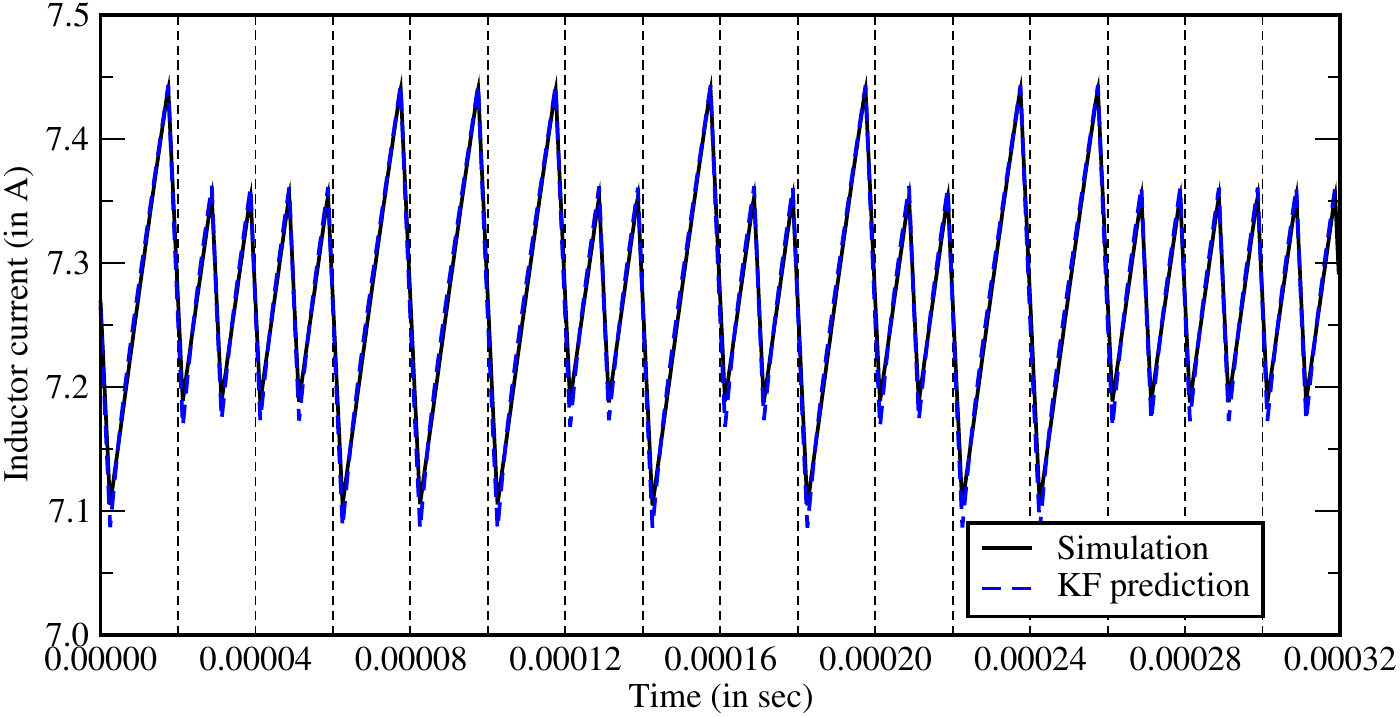}
\caption{Simulated and predicted inductor current.}
\end{subfigure}

\vspace{3mm}
\begin{subfigure}{0.485\textwidth}
\includegraphics[width=\linewidth]{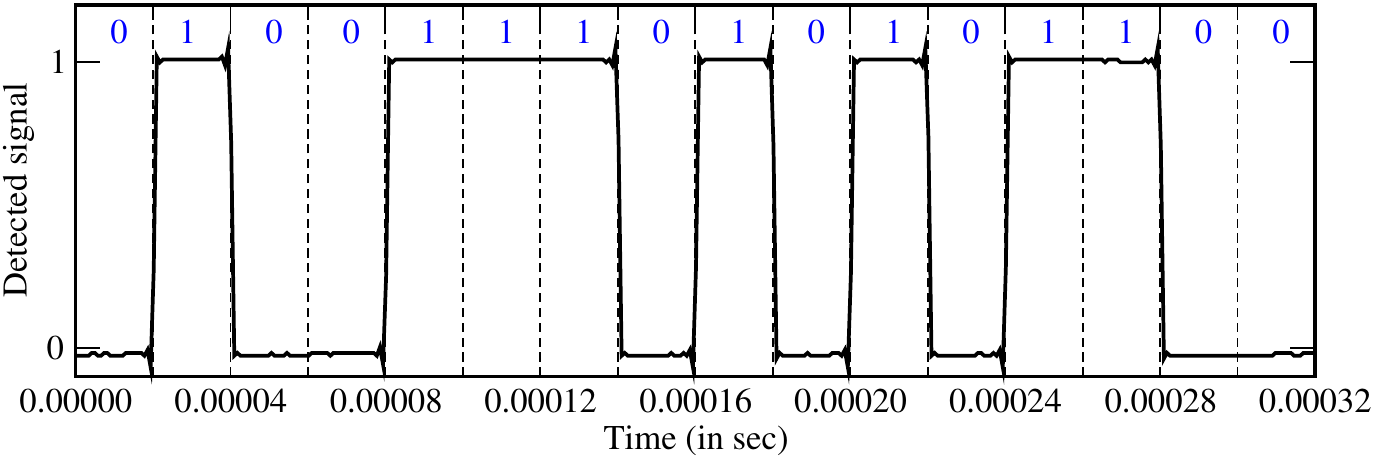}
\caption{Detected signal.}
\end{subfigure}

\vspace{3mm}
\begin{subfigure}{0.485\textwidth}
\includegraphics[width=\linewidth]{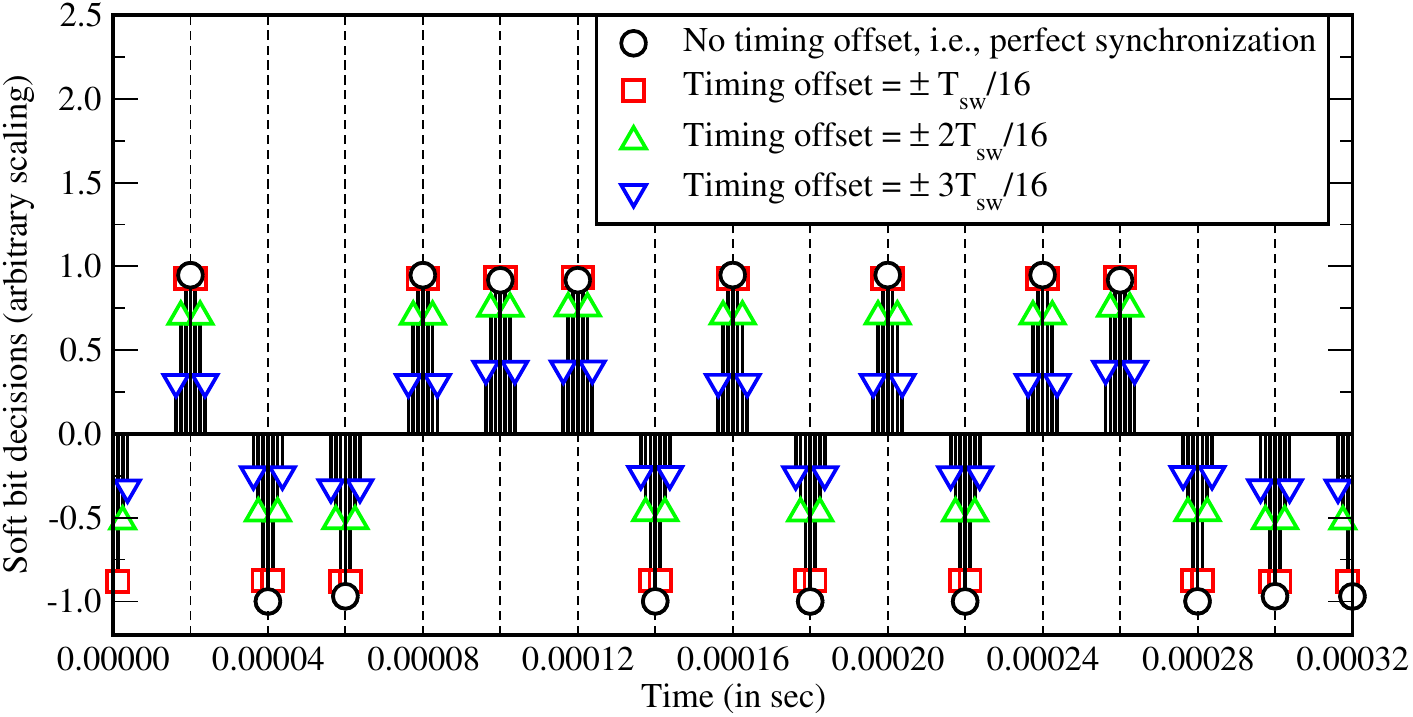}
\caption{Soft bit decisions for perfect and non-perfect timing sync.}
\end{subfigure}

\caption{Simulated and measured waveforms for Case Study I.}
\label{results1}
\end{figure}

\subsection{Case Study I: Bit Detection}
The purpose of this first case study is manifold: (i) For the first time, bit detection is demonstrated for a remote DT; (ii) It is shown that the discrete-time system modeling is applicable to Kalman filtering even for just 16 samples per symbol duration $T$; (iii) KF-assisted timing synchronization is investigated. 
For didactic reasons, this case study--unlike the other two case studies--examines the stationary case. 

First, the state-space model (\ref{x})-(\ref{yk}), which is based on the discrete-time model (\ref{1})-(\ref{2}) proposed in \cite{Hoeher2021}, was implemented in ANSI~C using $K_\mathrm{Tx}=1024$ samples per switching period $T_\mathrm{sw}$. 
Increasing $K_\mathrm{Tx}$ does not lead to any noteworthy improvement \cite{Mietzner2025}. 
The corresponding FSK-based switching signal and the output voltage of the PEC are plotted in Fig.~\ref{results1}\,(a) and (b), respectively. 
As expected, switching frequencies $f_0$ and $f_1$ cause different ripple voltages as shown by the black curve. 

Furthermore, in Fig.~\ref{results1}\,(b), corresponding experimental results are shown on the same time scale. 
In red color, measurements sampled by the digital oscilloscope are depicted. 
The close match between the analytical and measured waveforms--also reflected in the numerical results for the other case studies--confirms the fidelity of the experimental prototype, demonstrating very good agreement between ripple voltage simulation and experimental results. 
In blue color, the output voltage predicted by the KF is illustrated using data detection according to (\ref{hat_w}). 
As explained at the end of this subsection, the stationary Kalman gain is set to ${\bf K}=[K_1,K_2]=[0.5,0.8]^T$. 
In the KF, the state-space model was reproduced with just $K_\mathrm{Rx}=16$ samples per switching period in order to minimize computational complexity. 
The discrepancy between $K_\mathrm{Tx}=1024$ and $K_\mathrm{Rx}=16$ naturally leads to a modeling error.
The consistency with the simulated curve is still quite good. 
This finding indicates that the discrete-time model is subject to a sufficiently small modeling error even for a fairly large time step $\Delta T=T/K_\mathrm{Rx}$, and demonstrates the effectiveness of Kalman filtering for the DT application considered here. 
KF estimation results of the hidden inductor current are shown in Fig.~\ref{results1}\,(c) and compared with simulations of the discrete-time model. 
Again, the fit with the simulated curve is very good. 
Bit detection was error-free under lab conditions as long as timing synchronization was sufficiently precise. 
The latency between the output port of the half-bridge and the data output port of the STM32~H7 was measured as about 20~$\mu$s, i.\,e.~one switching period, compare Figs.~\ref{results1}\,(a) and (d). 

In Fig.~\ref{results1}\,(e) the soft bit decisions $\omega$ defined in (\ref{soft}) are plotted for 16 consecutive bit intervals. 
The sign corresponds to the hard bit decision, while the magnitude is an indicator for the bit-wise reliability. 
Interestingly, the latter can be used for timing synchronization. 
With increasing timing offset, the bit reliability gets worse as shown in the same figure. 

\begin{figure}[t]
\centering
\begin{subfigure}{0.485\textwidth}
\includegraphics[width=\linewidth]{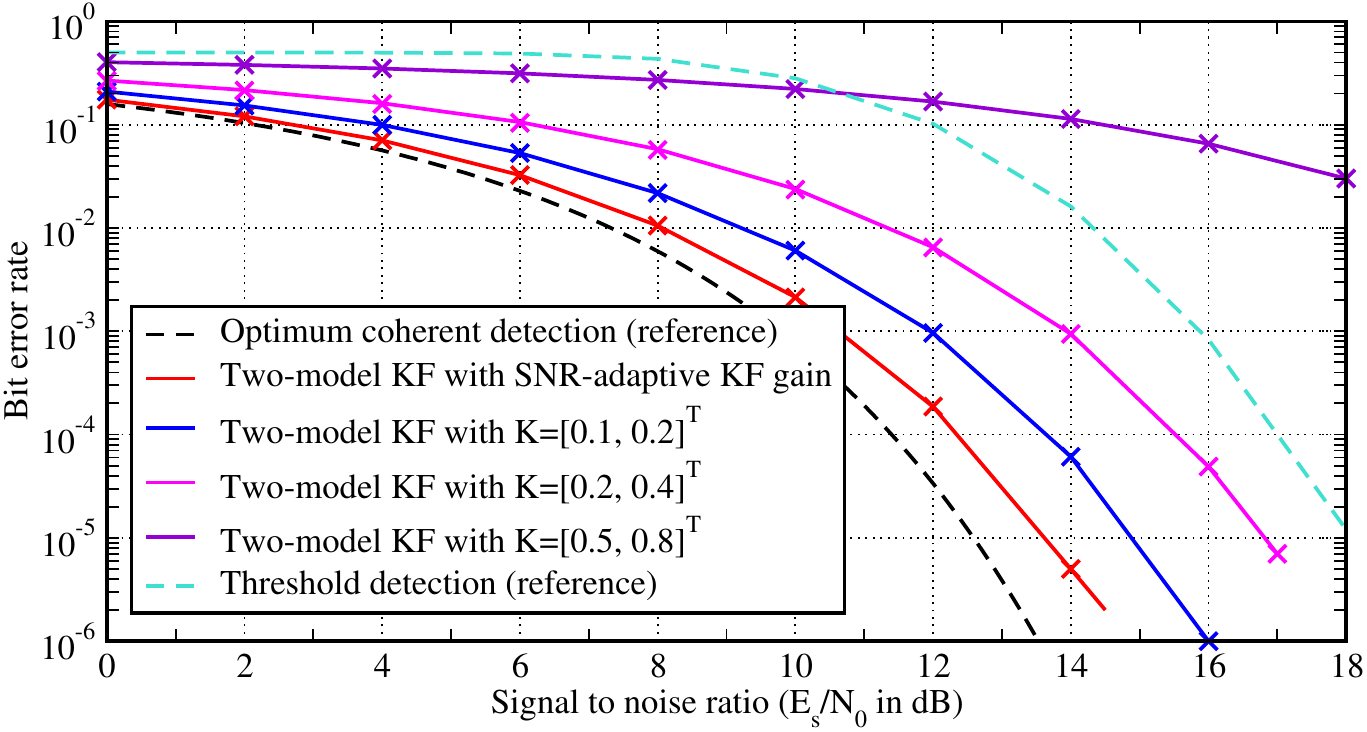}
\caption{Bit error rate.}
\end{subfigure}

\begin{subfigure}{0.485\textwidth}
\includegraphics[width=\linewidth]{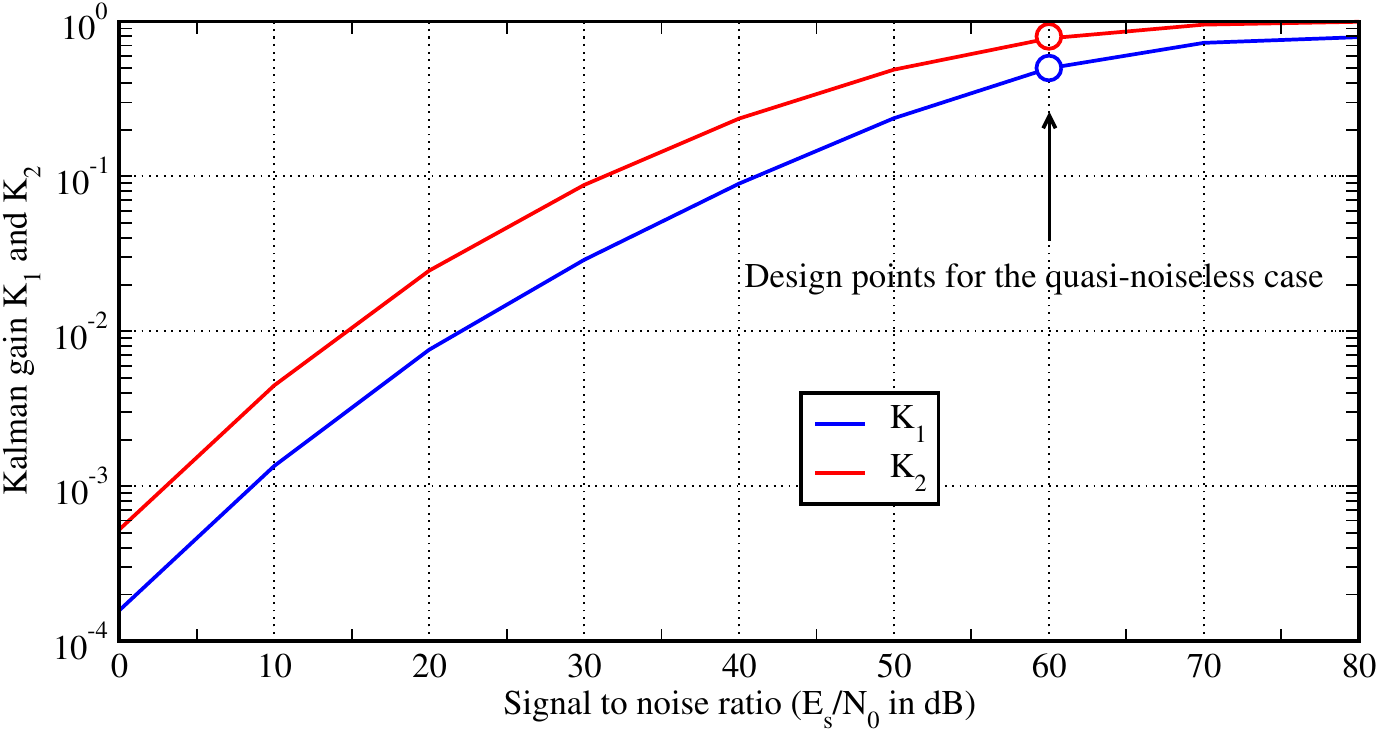}
\caption{Kalman filter gain.}
\end{subfigure}

\begin{subfigure}{0.485\textwidth}
\includegraphics[width=\linewidth]{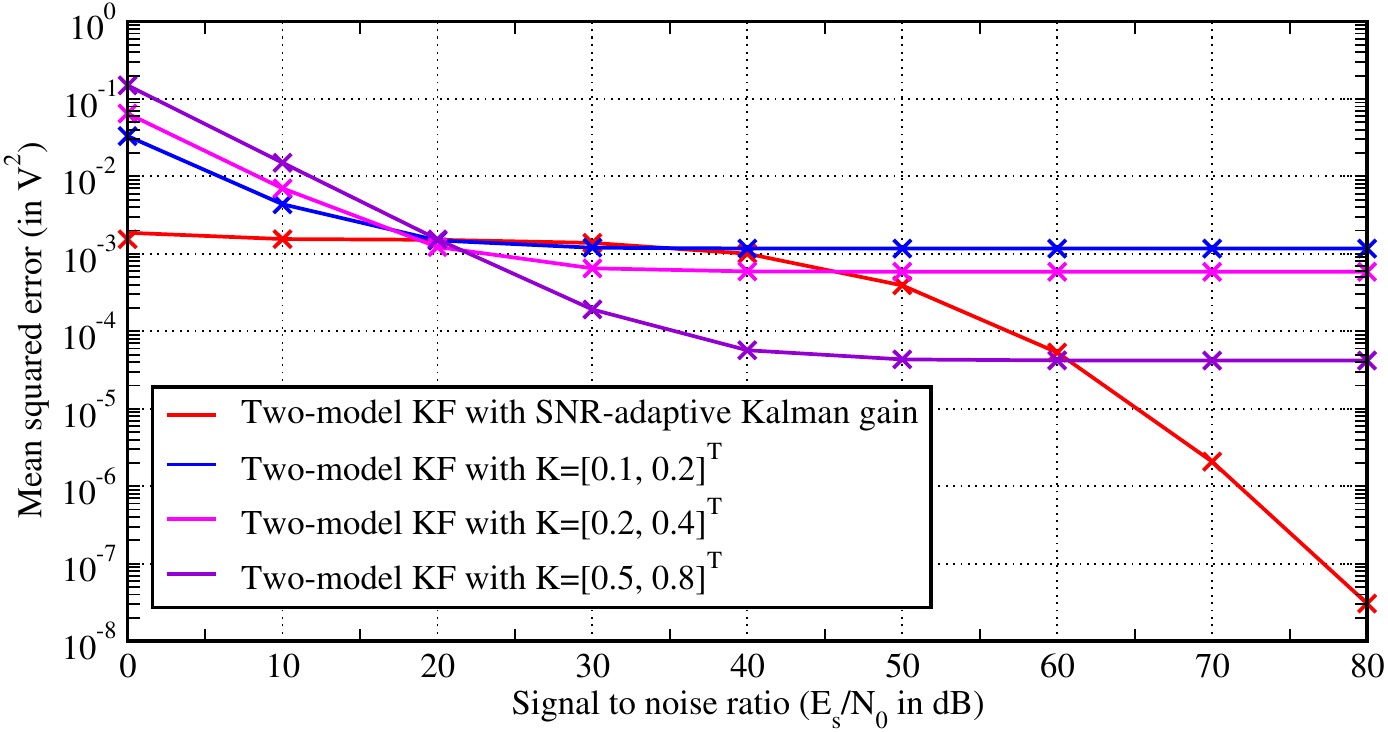}
\caption{Mean squared error.}
\end{subfigure}
\caption{Bit error rate (a), stationary Kalman filter gain (b), and mean squared error (c) versus signal-to-noise ratio simulation results for Case Study~I. In (a) and (c), the solid lines hold for hard model selection while the crosses belong to ideal model selection.} 
\label{fig:ber-snr}
\end{figure}

Since no bit errors were observed in the experiments and noise with scalable variance is difficult to inject in the analog domain, bit error simulations were additionally performed to assess the robustness of the proposed system. 
Towards this goal, additive white Gaussian noise with adjustable variance was added onto noiseless samples of the ripple voltage. 
The bit sequence defined above was repeated periodically. 
Corresponding bit error rate (BER) vs. SNR simulation results are depicted in Fig.~\ref{fig:ber-snr}\,(a). 
In order to eliminate the effect of $K_\mathrm{Rx}$ on the BER, on the x-axis $E_s/N_0$ is plotted in decibels, where $E_s$ is the energy per data symbol and $N_0/2$ is the double-sided noise spectral density \cite{Proakis2008}. 
In the binary case under investigation, the energy per bit $E_b$ and the energy per symbol $E_s$ are identical. 
The relation between SNR and the variance of the noise process, $\sigma_n^2=E\{n^2[k]\}$, is 
\begin{equation}
\sigma_n^2=\frac{K_\mathrm{Rx}\, \sigma_\mathrm{rip}^2}{2\, E_s/N_0}, 
\end{equation}
where $\sigma_\mathrm{rip}^2=E\{v_\mathrm{rip}^2\}$ is the ripple power. 
The ripple power has been determined from the waveform in Fig.~\ref{results1}\,(b) as $\sigma_\mathrm{rip}^2\approx 0.027~\mathrm{V}^2$. 
At $E_s/N_0=60$~dB, which is considered to be quasi-noiseless, $\sigma_n^2\approx 2.16\cdot 10^{-7}$. 
For the system under investigation, the measurement noise covariance ${\bf R}$ is scalar: ${\bf R}=\sigma_n^2$. 

Besides the noise covariance matrix ${\bf R}$, the Kalman gain is determined by the process-noise covariance matrix ${\bf Q}$. 
Ideally, for the two-model KF bank, ${\bf Q}={\bf 0}$, since the dynamical models are assumed to be deterministic. 
In practice, however, uncertainties are present in the dynamical process. 
These uncertainties may arise from the small modeling error discussed at the beginning of this subsection as well as from hardware imperfections. 
Therefore, the process-noise covariance matrix can be modeled as
\begin{equation}
{\bf Q} = 
\begin{bmatrix}
\left(\frac{\Delta T}{L}\right)^2 \sigma_v^2 & 0\\
0 & 0
\end{bmatrix},
\end{equation}
where $\sigma_v^2$ denotes the variance of an equivalent voltage disturbance acting on the inductor. 
We set $\sigma_v^2=2\cdot 10^{-2}~\mathrm{V}^2$. % which corresponds to about 141~mV RMS equivalent voltage-model error.
According to equations (\ref{A1})-(\ref{A3}) in the appendix, the corresponding stationary Kalman gain vector (which applies for this stationary case study) is ${\bf K}\approx [0.5,0.8]^T$ at 60~dB.
This configuration is used in the experiments of Case Studies I and II. 
Alternatively, the second diagonal element of ${\bf Q}$ can be modeled as $\left(\frac{\Delta T}{C}\right)^2 \sigma_i^2$, where $\sigma_i^2$ represents the variance of an equivalent current disturbance entering the capacitor equation.

Using a different set ($\sigma_v^2$, $\sigma_n^2$) as a starting point would certainly change the Kalman gain, but it is interesting to mention that scaling both $\sigma_v^2$ and $\sigma_n^2$ by the same factor leaves the Kalman gain unchanged in the steady state. 
As illustrated in Fig.~\ref{fig:ber-snr}\,(b), the stationary Kalman gain strongly depends on the SNR if $\sigma_v^2$ is fixed (here: $\sigma_v^2=2\cdot 10^{-2}~\mathrm{V}^2$). 
For an SNR-adaptive stationary Kalman gain, the BER represented by the red curve in Fig.~\ref{fig:ber-snr},(a) improves significantly over that achieved with ${\bf K}=[0.5,0.8]^T$.
Compared to an optimal coherent FSK detector \cite{Proakis2008}, which serves as a reference and is represented by the black dashed lines, even in the low-SNR range the loss is only 1 dB (which is largely attributable to ISI).
Besides ${\bf K}=[0.5,0.8]^T$, in Fig.~\ref{fig:ber-snr}\,(a) the BER is plotted for several other fixed Kalman gain vectors as well, all taken from Fig.~\ref{fig:ber-snr}\,(b). 
For fixed Kalman gain vectors the loss is generally attributable largely to non-perfect channel estimation. 
In practice, however, a fixed but suitable Kalman gain should suffice, since SNRs below 20~dB are unlikely to occur.
For the purpose of benchmarking, in Fig.~\ref{fig:ber-snr}\,(a) BER results are also included for a conventional threshold detector. 
The threshold detector calculates the absolute values of the noisy ripple voltage, sums them over one symbol period (i.e., over $K_\mathrm{Rx}$ samples), and compares the sum to a threshold. 
The computational complexity of this threshold detector is on the same order as a 2-state stationary KF. 

In Fig.~\ref{fig:ber-snr}(c), the corresponding mean squared error (MSE) curves are plotted.
The MSE is computed as the mean squared difference between the ripple voltage estimated by the KF and the noiseless ripple voltage.
Note that an MSE of $10^{-3}~\mathrm{V}^2$ corresponds to 3.7~\% of the ripple power, a smaller MSE would be of little benefit. 

So far, hard model selection as defined above was performed, i.\,e., the actual bit decision determines the KF estimates for the next bit interval. 
This selection process may cause error propagation. 
For this reason, in additional computer simulations a genie-aided receiver was tested with ideal model selection, meaning that it always selects the correct bit.  
Results are shown by crosses (``x'') in Fig.~\ref{fig:ber-snr}\,(a) and Fig.~\ref{fig:ber-snr}\,(c). 
The fact that no significant discrepancies can be observed even in the low-noise region is an indicator of the method's robustness. 

\begin{figure}[t]
\centering
\begin{subfigure}{0.485\textwidth}
\includegraphics[width=\linewidth]{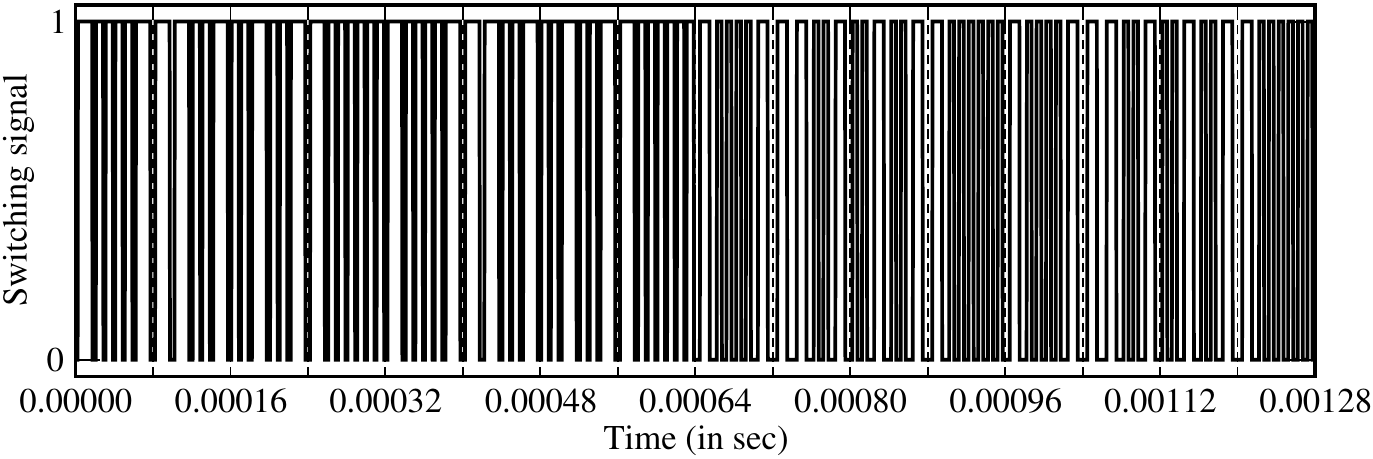}
\caption{FSK-modulated switching signal.}
\end{subfigure}

\vspace{3mm}
\begin{subfigure}{0.485\textwidth}
\includegraphics[width=\linewidth]{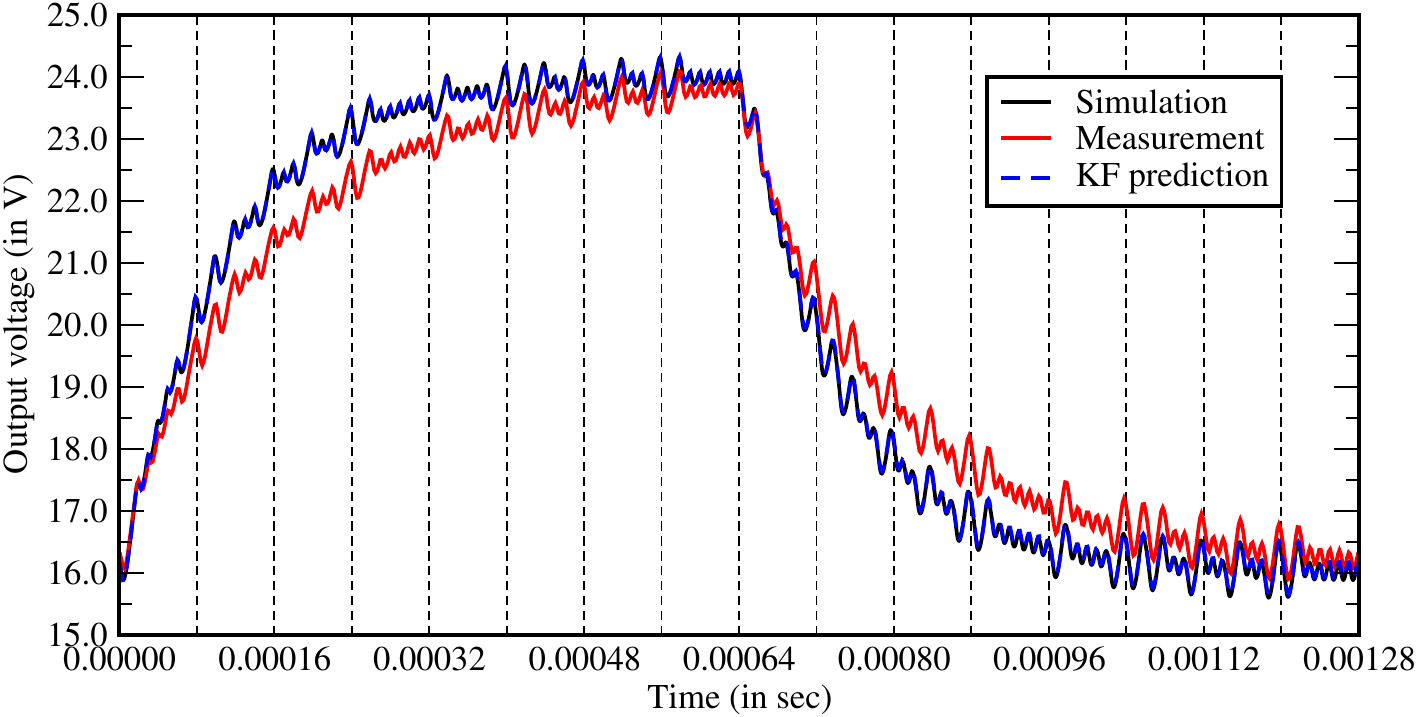}
\caption{Simulated and measured output voltage.}
\end{subfigure}

\vspace{3mm}
\begin{subfigure}{0.485\textwidth}
\includegraphics[width=\linewidth]{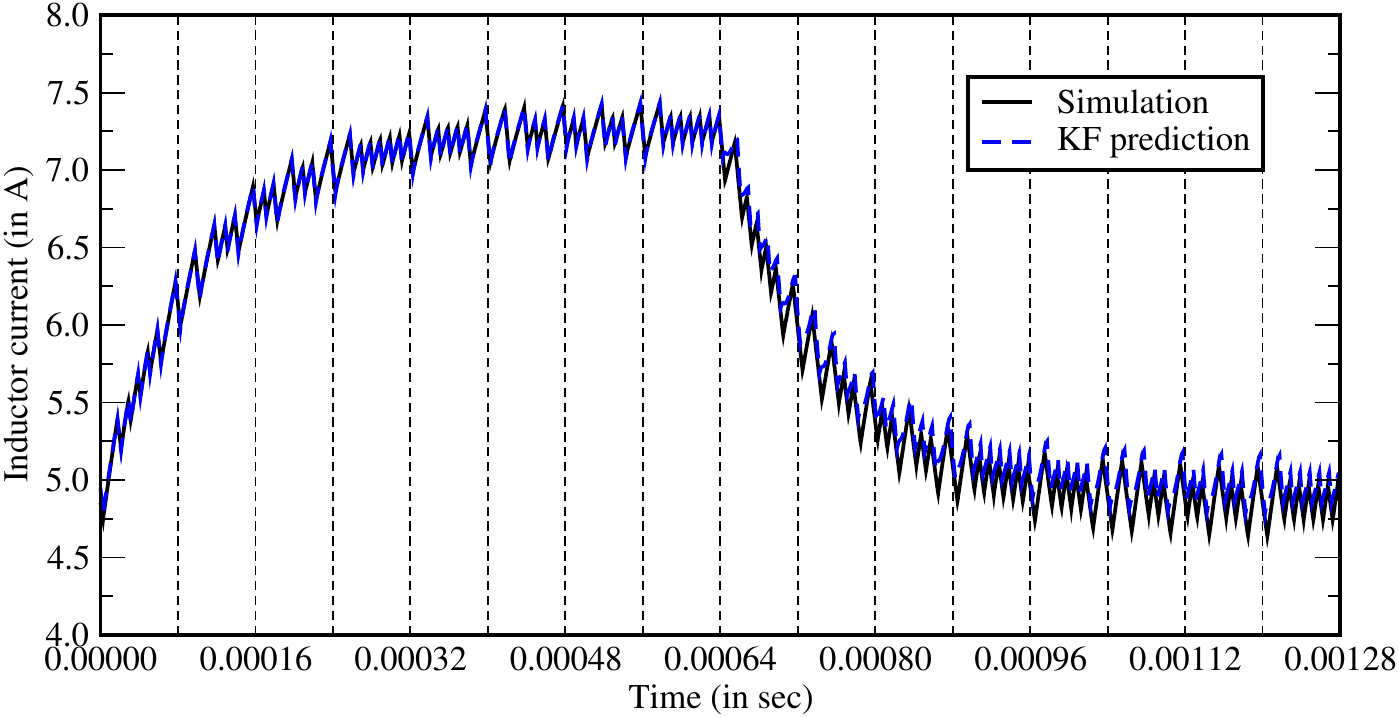}
\caption{Simulated and predicted inductor current.}
\end{subfigure}

\caption{Simulated and measured waveforms for Case Study II.}
\label{results2}
\end{figure}

\subsection{Case Study II: Adaptation to Output Voltage Variations}
The purpose of the second case study is to verify that a KF-based DT is able to precisely react on abrupt output voltage variations. 
These are triggered by periodically toggling between duty cycles $D_1=0.75$ and $D_2=0.5$, see Fig.~\ref{results2}\,(a). 
Consequently, the average output voltage is 20~V, while the steady-state voltages are 24~V and 16~V, respectively. 
The transient step response lasts about $300~\mu$s.
The stationary Kalman gain is again set to ${\bf K}=[0.5,0.8]^T$. 
As depicted in Figs.~\ref{results2}\,(b) and (c), the KF-based DT is able to track the transient behavior quite well for the same fixed Kalman gain used in Case Study~I.  
The color code is adopted from Figs.~\ref{results1}\,(b) and (c). 
FSK-based data transfer can be carried out seamlessly, even in the worst case of a step-wise duty-cycle variation. 

\begin{figure}[t]
\centering
\begin{subfigure}{0.485\textwidth}
\includegraphics[width=\linewidth]{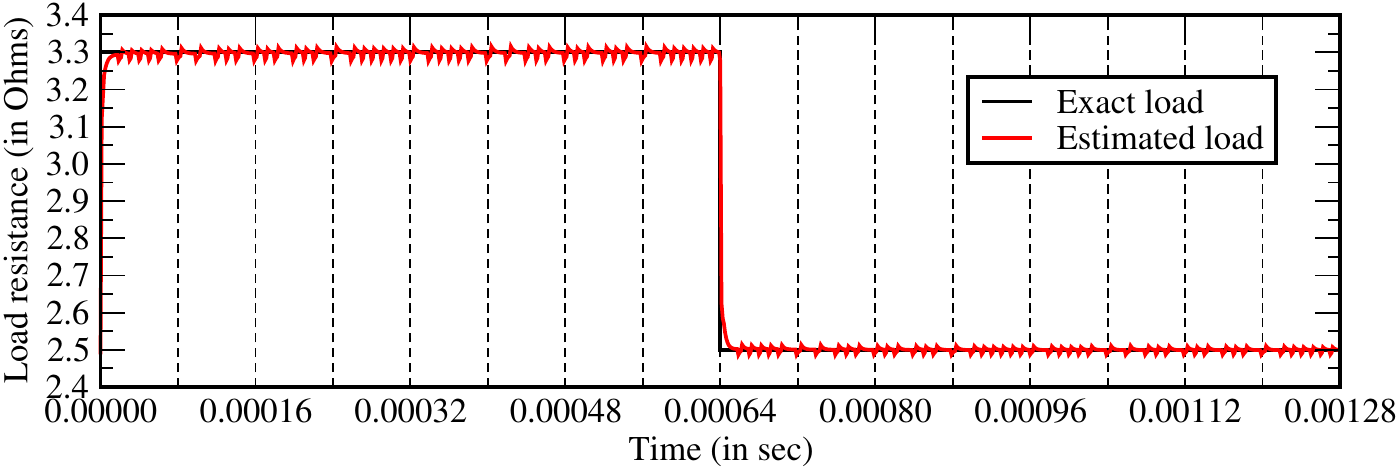}
\caption{Exact and simulated load resistance.}
\end{subfigure}

\vspace{3mm}
\begin{subfigure}{0.485\textwidth}
\includegraphics[width=\linewidth]{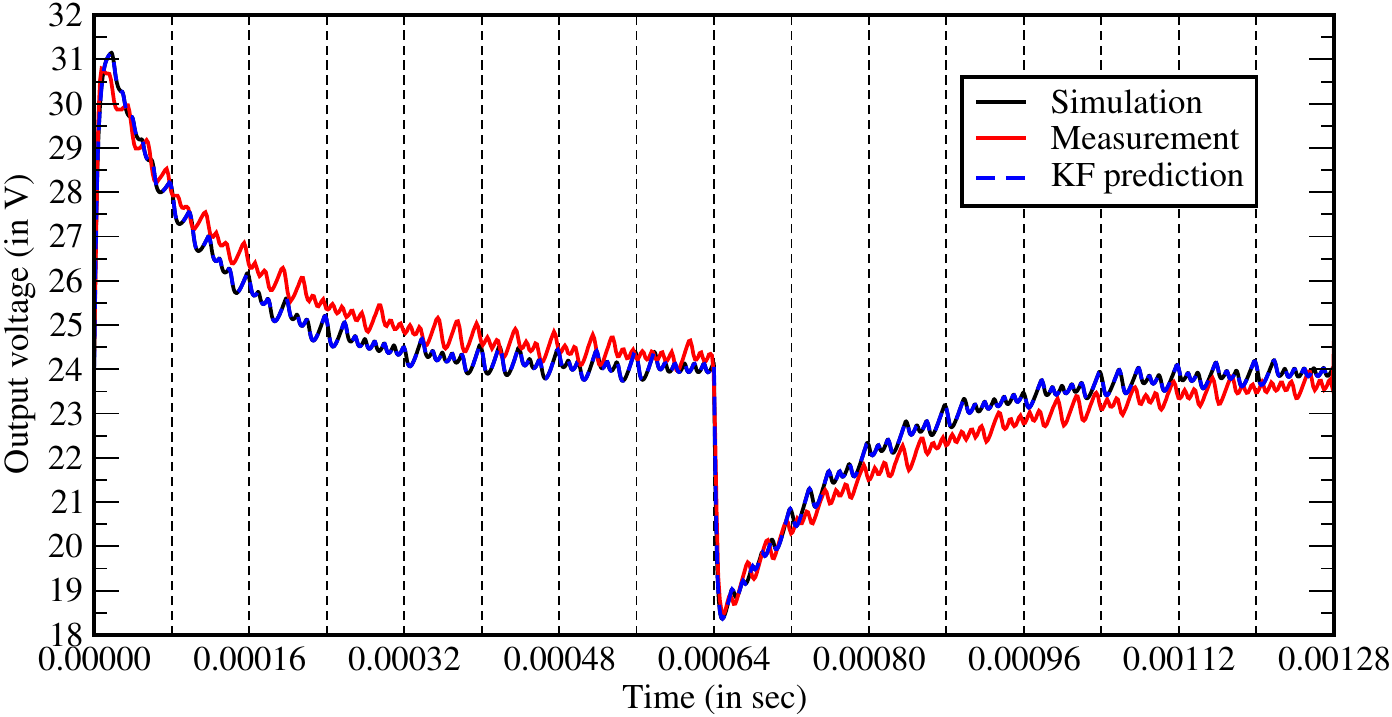}
\caption{Simulated and measured output voltage.}
\end{subfigure}

\vspace{3mm}
\begin{subfigure}{0.485\textwidth}
\includegraphics[width=\linewidth]{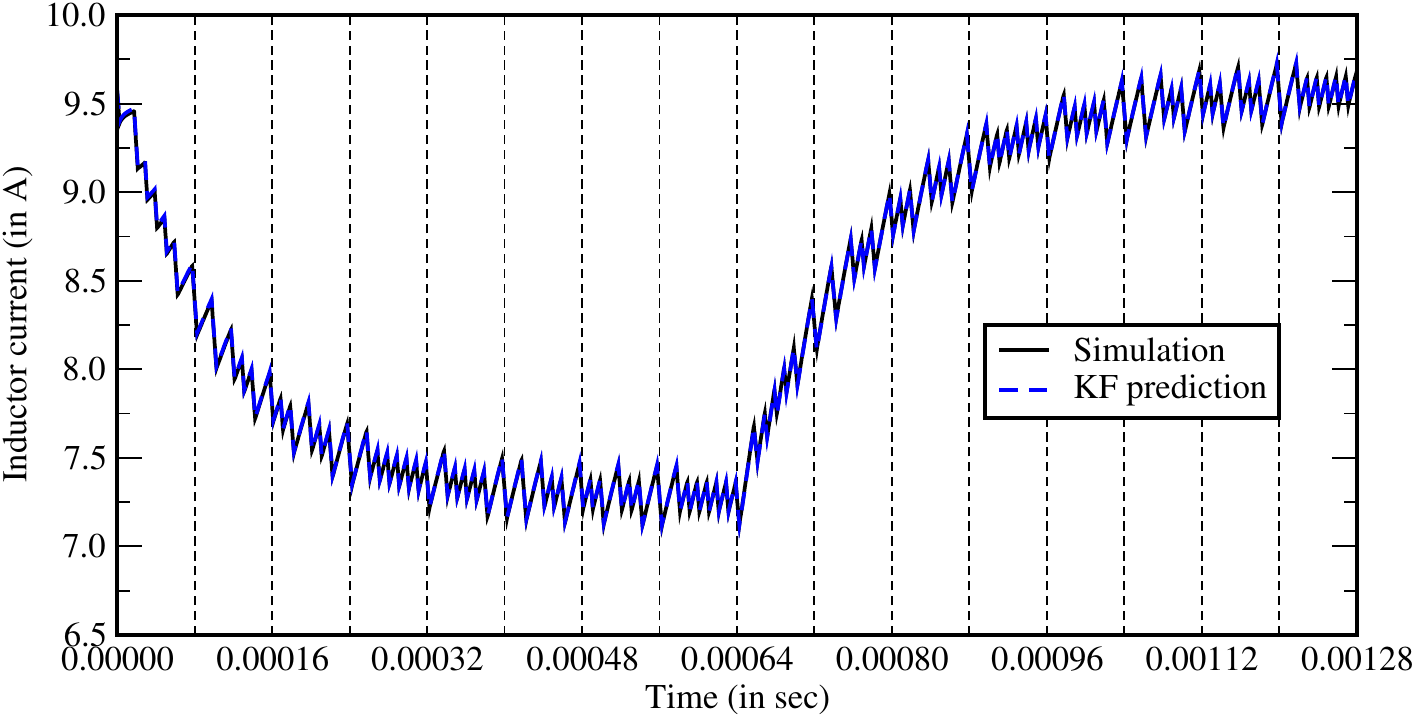}
\caption{Simulated and predicted inductor current.}
\end{subfigure}

\caption{Simulated and measured waveforms for Case Study III.}
\label{results3}
\end{figure}

\subsection{Case Study III: Estimation of Load Changes}
The purpose of the third case study is to verify that a KF-based DT is able to precisely react on abrupt load changes. 
Towards this goal, the load is periodically toggling between $R_1=3.3~\Omega$ and $R_2=2.5~\Omega$, see Fig.~\ref{results3}\,(a). 
As mentioned above, load estimation is performed by means of an AKF in combination with hard model selection. 
Unlike Case Studies I and II, a time-varying Kalman gain is used here, which is calculated according to equations (\ref{App1})–(\ref{App6}) in the appendix using the $3\times 3$ process-noise covariance matrix 
\begin{equation}
{\bf Q} = 
\begin{bmatrix}
\left(\frac{\Delta T}{L}\right)^2 \sigma_v^2 & 0 & 0\\
0 & \left(\frac{\Delta T}{C}\right)^2 \sigma_i^2 & 0\\
0 & 0 & \sigma_g^2
\end{bmatrix}
\end{equation}
and the noise variance ${\bf R}=\sigma_n^2$. 
Like above, we set $\sigma_n^2=2.16\cdot 10^{-7}$, $\sigma_v^2=0.02$, $\sigma_i^2=0$, and additionally $\sigma_g^2=E\{q_g^2[k]\}=0.01$. 
As shown in Fig.~\ref{results3}\,(a), the AKF is able to track the fast transient behavior very well. 
The influence of the data-dependent ripple voltage on the parameter estimation is small. 
Voltage and current state-estimation on the one hand and load estimation on the other hand can be performed simultaneously without significant degradation, as illustrated in Figs.~\ref{results3}\,(b) and (c). 

\section{Conclusion}\label{sec:conclusion} 
Integrating a talkative power electronics converter--capable of embedding information into its ripple voltage--with a digital twin based on state-space modeling in combination with Kalman filtering is conceptually compelling, as it allows the converter to communicate its internal states over a certain distance, which depends on noise and channel characteristics. 
Corresponding information bits can be reliably decoded and used to continuously update the real-time virtual model. 
This approach effectively provides a methodology for monitoring, diagnostics, and predictive control, enabling a tight coupling between spatially-distributed physical and digital domains while avoiding external communication infrastructure. 
Although challenges such as limited bandwidth and the trade-off between power quality and bit reliability must be managed, the combination represents a promising direction for cyber-physical energy systems in which power electronics converters become more adaptive and intelligent. 
The feasibility of the proposed concept is verified for three selected case studies: bit detection, adaptation to output voltage variations, and estimation of load variations. 
All three case studies are numerically supported by simulation results (obtained with ANSI~C code) and by lab-based experimental results (involving online Kalman filter computations on an STM32~H7 microcontroller and offline processing of the sampled ripple voltage).
Depending on the task, namely bit detection or parameter estimation, either a multi-model Kalman filter bank is applied or the states of a conventional Kalman filter are augmented for parameter estimation. 

\section*{Appendix: Kalman Filter}
A discrete-time KF is, for a given discrete-time state-space model, the optimal LMMSE estimator of the state vector ${\bf x}[k]$, based on the measurements available up to time index $k$. In particular, the KF minimizes the mean-square estimation error, and therefore minimizes the error covariance matrix $\mbox{\boldmath $\Sigma$}[k|k]$ in the positive-semidefinite sense. Consequently, it also minimizes the trace of the error covariance matrix, $\mathrm{tr}\{\mbox{\boldmath $\Sigma$}[k|k]\}$.
For the system under investigation, the calculation steps are \cite{Anderson1979, Sarkka2023}: 
\begin{enumerate}
\item State prediction (see Fig.~\ref{fig:kalman}): 
\begin{equation}\label{App1}
\hat {\bf x}[k+1|k] = {\bf A}\, \hat {\bf x}[k|k-1] + {\bf B}\, u[k] + {\bf K}[k]\, \big(y[k]-\hat y[k]\big)
\end{equation}
\item Measurement equation (see Fig.~\ref{fig:kalman}): 
\begin{equation} \label{App2}
\hat y[k] = {\bf C}^T\, \hat {\bf x}[k|k-1]
\end{equation}
\item Error covariance matrix:
\begin{align}\label{App3} 
\mbox{\boldmath $\Sigma$}[k&|k-1] =\nonumber \\
& E\left\{({\bf x}[k] - \hat{\bf x}[k|k-1]) ({\bf x}[k] - \hat{\bf x}[k|k-1])^T\right\}
\end{align}
\item Kalman gain:
\begin{equation}\label{App4}
{\bf K}[k] = \mbox{\boldmath $\Sigma$}[k|k-1]\, {\bf C}\, \Big({\bf C}^T\, 
\mbox{\boldmath $\Sigma$}[k|k-1]\, {\bf C} + {\bf R} \Big)^{-1}
\end{equation}
\item Measurement update:
\begin{equation}\label{App5} 
\mbox{\boldmath $\Sigma$}[k|k] = \mbox{\boldmath $\Sigma$}[k|k-1] - {\bf K}[k]\, {\bf C}^T\, \mbox{\boldmath $\Sigma$}[k|k-1]
\end{equation}
\item Time update:
\begin{equation}\label{App6} 
\mbox{\boldmath $\Sigma$}[k+1|k] = {\bf A}\, \mbox{\boldmath $\Sigma$}[k|k]\, {\bf A}^T + {\bf Q}. % {\bf B} {\bf Q} {\bf B}^T.
\end{equation}
%% \item {\bf Initialisierung}:
%% $$\hat {\bf x}[0|-1] = \bar {\bf x}[0]; ~\mbox{\boldmath $\Sigma$}[0|-1] = E\left\{ [{\bf x}[0]-\bar {\bf x}[0]]\cdot 
%% [{\bf x}[0]-\bar {\bf x}[0]]^T\right\}.$$
\end{enumerate}
Note that the KF is a copy of the original system, see Fig.~\ref{fig:kalman}, driven by the innovation process $e[k]=y[k]-\hat y[k]$. 
If the system and noise covariance matrices ${\bf A}$, ${\bf C}$, ${\bf Q}$, and ${\bf R}$ are time-invariant and the usual conditions for convergence of the Riccati recursion are satisfied, the error covariance and Kalman gain converge to steady-state values. 
Then, the stationary Riccati equation is yielded as 
\begin{align}\label{A1}
\mbox{\boldmath $\Sigma$} &= \lim_{k\to\infty} \mbox{\boldmath $\Sigma$}[k|k-1] \stackrel{2\times 2}{=} 
\begin{bmatrix}
\sigma_{1}^2 & \sigma_{12} \\
\sigma_{21} & \sigma_{2}^2
\end{bmatrix}
\\
&= {\bf A}\Big[\mbox{\boldmath $\Sigma$} - \mbox{\boldmath $\Sigma$}\, {\bf C}\Big({\bf C}^T\, \mbox{\boldmath $\Sigma$}\, {\bf C} + \sigma_n^2\Big)^{-1}\, {\bf C}^T\, \mbox{\boldmath $\Sigma$}\Big]{\bf A}^{T} + {\bf Q} \nonumber % {\bf B} {\bf Q} {\bf B}^T \nonumber
\end{align}
and the stationary Kalman gain vector is 
\begin{equation}\label{A2}
{\bf K} = \lim\limits_{k\to\infty} {\bf K}[k] = 
\mbox{\boldmath $\Sigma$}\, {\bf C}\Big({\bf C}^{T}\, \mbox{\boldmath $\Sigma$}\, {\bf C} + \sigma_n^2\Big)^{-1}. 
\end{equation}
For a $2\times 2$ system with ${\bf C}=[0, 1]^T$, the gain vector simplifies as
\begin{equation}\label{A3}
{\bf K} = [\sigma_{12}, \sigma_{22}]^T / (\sigma_{2}^2+\sigma_n^2).
\end{equation}

\bibliographystyle{IEEEtran}
\bibliography{IEEEabrv, references}

@article{Stefanutti2008,
  title={Power line communication in digitally controlled {DC-DC} converters using switching frequency modulation},
  author={Stefanutti, Walter and Saggini, Stefano and Mattavelli, Paolo and Ghioni, Massimo},
  journal={IEEE Trans. Ind. Electron.},
  volume={55},
  number={4},
  pages={1509-1518},
  year={2008},
}

@article{Wu2015,
  title={Power conversion and signal transmission integration method based on dual modulation of {DC-DC} converters},
  author={Wu, Jiande and Du, Jin and Lin, Zhengyu and Hu, Yihua and Zhao, Chongwen and He, Xiangning},
  journal={IEEE Trans. Ind. Electron.},
  volume={62},
  number={2},
  pages={1291-1300},
  year={2015},
}

@article{Popovski2016,
  title={Multiuser communication through power talk in {DC} microgrids},
  author={Angjelichinoski, Marko and Stefanovi{\'c}, {\v{C}}edomir and Popovski, Petar and Liu, Hongpeng and Loh, Poh Chiang and Blaabjerg, Frede},
  journal={IEEE J. Sel. Areas Commun.},
  volume={34},
  number={7},
  pages={2006-2021},
  year={2016},
  publisher={IEEE}
}

@article{He2020,
  title={Nature of power electronics and integration of power conversion with communication for talkative power},
  author={He, Xiangning and Wang, Ruichi and Wu, Jiande and Li, Wuhua},
  journal={Nat. Commun.},
  volume={11},
  number={1},
  pages={1-12},
  year={2020},
}

@article{Hoeher2021,
  title={Channel coding and receiver design for simultaneous wireline information and power transfer},
  author={Hoeher, Peter A. and Mewis, Maximilian and Liserre, Marco},
  journal={IEEE Open J. Power Electron.},
  volume={2},
  pages={545-558},
  year={2021},
}

@article{Wang2023,
  author={Wang, Ruichi and He, Xiangning and Wu, Jiande and Zhang, Ruoqi and Li, Wuhua},
  journal={IEEE J. Emerg. Sel. Top. Power Electron.}, 
  title={Power and signal dual modulation with info nature of power converters}, 
  year={2023},
  volume={11},
  number={1},
  pages={588-601},
  doi={10.1109/JESTPE.2022.3199099}
}

@article{Liserre2023,
  title={Overview of talkative power conversion technologies},
  author={Liserre, Marco and Beiranvand, Hamzeh and Leng, Yang and Zhu, Rongwu and Hoeher, Peter A.},
  journal={IEEE Open J. Power Electron.},
  volume={4},
  pages={67-80},
  year={2023},
  publisher={IEEE}
}

@article{Moench2023,
  title={Efficient talkative power conversion with quasi-square-wave zero-voltage switching hysteretic current control},
  author={Moench, Stefan and Kuring, Carsten and Geng, Xiaomeng and Hoeher, Peter A. and Liserre, Marco and Quay, Rudiger and Dieckerhoff, Sibylle},
  journal={IEEE Trans. Power Electron.},
  volume={38}, 
  number={10},
  pages={12383-12387},
  year={2023},
  publisher={IEEE}
}

@article{Mousavi2024,
  title={Empowering talkative power technology in wireless power transfer with machine learning},
  author={Mousavi, Seyed Ali and Ghahramanlzadi, Zahra and Khooban, Mohammad Hassan},
  journal={IET Power Electron.},
  volume={17},
  number={16},
  pages={3083-3092},
  year={2024},
  publisher={Wiley Online Library}
}

@article{PIEEE2025,
  title={Talkative power conversion: A tutorial},
  author={Hoeher, Peter A. and Leng, Yang and Zhu, Rongwu and Liserre, Marco},
  journal={Proc. IEEE},
  volume={113},
  number={4},
  pages={344-369},
  year={2025}
}

@article{Mietzner2025,
  title={Comprehensive Signal Modeling for Talkative Power Conversion},
  author={Mietzner, Jan and Chakraborty, Cerikh and Hoeher, Peter A. and Lampe, Lutz},
  journal={IEEE Access},
  volume={13},
  pages={111098--111114}, 
  year={2025},
  publisher={IEEE}
}

@article{Hoeher2025multiport,
  title={Multiport Talkative Power Conversion: Basic Principle and System Modeling},
  author={Hoeher, Peter A. and Leng, Yang and Zhu, Rongwu},
  journal={IEEE Open J. Power Electron.},
  volume={6},
  pages={1174--1189},
  year={2025},
  publisher={IEEE}
}

@article{Hoeher2025Shannon,
  title={Information Theoretical Bounds for Talkative Power Conversion},
  author={Hoeher, Peter A. and Beiranvand, Hamzeh and Liserre, Marco},
  journal={IEEE Trans. Smart Grids},
  volume={16},
  number={6},
  pages={5655-5668},
  year={2025},
  publisher={IEEE}
}

@article{Mewis2026a,
  author={Mewis, Maximilian and Hoeher, Peter A.},
  title={Cyber-Physical System Aspects of Talkative Power Conversion Networks},
  journal={IEEE Access},
  volume={14},
  number={},
  pages={5002-5011},
  year={2026},
}

@article{Yoon2026,
  title={Embedded {AES} Encryption for Secure {FSK}-Based Talkative Power Conversion With {DC-DC} Converters},
  author={Yoon, Geun-Ho and Kim, Mina and Park, Hwa-Pyeong},
  journal={IEEE Trans. Power Electron.},
  volume={41},
  number={6},
  pages={ 9468-9479},
  year={2026},
  publisher={IEEE}
}

@article{Hoeher2026a,
  title={{MIMO}-enhanced talkative power conversion: A perspective},
  author={Hoeher, Peter A. and Leng, Yang and Zhu, Rongwu},
  journal={Commun. Eng.},
  volume={5},
  number={131},
  pages={},
  year={2026}, 
  publisher={Nature}
}

@article{Hoeher2026b,
  author={Hoeher, P. A. and Leng, Y. and Mewis-Noll, M. and Zhu, R.}, 
  title={Power line communication vs. talkative power conversion: Benchmarking and hybrid combination},
  journal={IEEE Ind. Electron. Mag.}, 
  year={(Early Access), 2026},
}

@inproceedings{Chakraborty2026,
  author={Chakraborty, C. and Lampe, L. and Mietzner, J. and Hoeher, Peter A.},
  title={Talkative power conversion: Tackling challenging channels via reduced-state equalization},
  booktitle={Proc. IEEE ICC '26, Glasgow, Scotland, UK},
  pages={1-6}, 
  year={2026},
}

@article{Mewis2026b,
  author={Mewis-Noll, M. and Hoeher, Peter A.}, 
  title={Cost-effective implementation of talkative power conversion utilizing software-defined modems},
  journal={IEEE Open J. Power Electron.}, 
  year={(Early Access), 2026},
}

@article{Tao2018,
  title={Digital twin in industry: State-of-the-art},
  author={Tao, Fei and Zhang, He and Liu, Ang and Nee, Andrew YC},
  journal={IEEE Trans. Ind. Inform.},
  volume={15},
  number={4},
  pages={2405--2415},
  year={2018},
  publisher={IEEE}
}

@article{Tao2024,
  title={Advancements and challenges of digital twins in industry},
  author={Tao, Fei and Zhang, He and Zhang, Chenyuan},
  journal={Nat. Comput. Sci.},
  volume={4},
  number={3},
  pages={169--177},
  year={2024},
  publisher={Nature Publishing Group US New York}
}

@article{Jain2020,
  title={A digital twin approach for fault diagnosis in distributed photovoltaic systems},
  author={Jain, Palak and Poon, Jason and Singh, Jai Prakash and Spanos, Costas and Sanders, Seth R and Panda, Sanjib Kumar},
  journal={IEEE Trans. Power Electron.},
  volume={35},
  number={1},
  pages={940--956},
  year={2020},
  publisher={IEEE}
}

@article{Peng2021,
  title={A digital twin based estimation method for health indicators of {DC--DC} converters},
  author={Peng, Yingzhou and Zhao, Shuai and Wang, Huai},
  journal={IEEE Trans. Power Electron.},
  volume={36},
  number={2},
  pages={2105--2118},
  year={2021},
  publisher={IEEE}
}

@article{Milton2020,
  title={Controller-embeddable probabilistic real-time digital twins for power electronic converter diagnostics},
  author={Milton, Matthew and De La O, Castulo and Ginn, Herbert L and Benigni, Andrea},
  journal={IEEE Trans. Power Electron.},
  volume={35},
  number={9},
  pages={9850--9864},
  year={2020},
  publisher={IEEE}
}

@article{Chen2023a,
  title={Digital twin techniques for power electronics-based energy conversion systems: A survey of concepts, application scenarios, future challenges, and trends},
  author={Chen, Haoyu and Zhang, Zhenbin and Karamanakos, Petros and Rodriguez, Jose},
  journal={IEEE Ind. Electron. Mag.},
  volume={17},
  number={2},
  pages={20--36},
  year={2023},
  publisher={IEEE}
}

@article{Bai2024,
  title={Digital twins for modern power electronics: An investigation into the enabling technologies and applications},
  author={Bai, Hao and Kuprat, Johannes and Os{\'o}rio, Caio and Liu, Chen and Liserre, Marco and Gao, Fei},
  journal={IEEE Electrif. Mag.},
  volume={12},
  number={3},
  pages={50--67},
  year={2024},
  publisher={IEEE}
}

@article{DiNezio2023,
  title={{DC-DC} boost converters parameters estimation based on digital twin},
  author={Di Nezio, Giulia and Di Benedetto, Marco and Lidozzi, Alessandro and Solero, Luca},
  journal={IEEE Trans. Ind. Appl.},
  volume={59},
  number={5},
  pages={6232--6241},
  year={2023},
  publisher={IEEE}
}

@article{Lei2023,
  title={Digital twin based monitoring and control for {DC-DC} converters},
  author={Lei, Zhongcheng and Zhou, Hong and Dai, Xiaoran and Hu, Wenshan and Liu, Guo-Ping},
  journal={Nat. Commun.},
  volume={14},
  number={1},
  pages={5604},
  year={2023},
  publisher={Nature Publishing Group}
}

@article{Wu2025,
  title={An overview of digital twin technology for power electronics: State-of-the-art and future trends},
  author={Wu, Chenhao and Cui, Zhexin and Xia, Qian and Yue, Jiguang and Lyu, Feng},
  journal={IEEE Trans. Power Electron.},
  volume={40},
  number={9},
  pages={13337-13362},
  year={2025},
  publisher={IEEE}
}

@article{Bhoi2025,
  title={Overview of Digital Twin Development in Power Electronics},
  author={Bhoi, Sachin Kumar and Chakraborty, Sajib and Hosseinabadi, Farzad and Frikha, Mohamed Amine and Martin, Gamze Egin and Sorniotti, Aldo and Hegazy, Omar},
  journal={IEEE Open J. Power Electron.},
  volume={6}, 
  pages={1737-1758}, 
  year={2025},
  publisher={IEEE}
}

@article{Zheng2025,
  title={Cognitive digital twins-based model predictive control for high-frequency power converters},
  author={Zheng, Jialin and Wang, Haoyu and Zeng, Yangbin and Xu, Han and Mou, Di and Li, Hong and Rodriguez, Jose},
  journal={IEEE Trans. Ind. Electron.},
  volume={72},
  number={12},
  pages={13310-13321},
  year={2025},
  publisher={IEEE}
}

@article{Barreto2025,
  title={Cyber-Physical Power System Digital Twins -- A Study on the State of the Art},
  author={Barreto, Nathan Elias Maruch and Aoki, Alexandre Rasi},
  journal={Energies},
  volume={18},
  number={22},
  pages={5960},
  year={2025},
  publisher={MDPI}
}

@article{Blaabjerg2026,
  title={Advances in Reliability and Artificial Intelligence for Power Electronic Systems},
  author={Blaabjerg, Frede and Zhang, Kaichen and Song, Yubo and Zhang, Yichi and Yao, Bo and Novak, Mateja and Zhao, Shuai and Sahoo, Subham and Sangwongwanich, Ariya and Wang, Huai},
  journal={IEEE Trans. Power Electron.},
  volume={41},
  number={8},
  pages={12386-12409},
  year={2026},
  publisher={IEEE}
}

@article{Notash2026,
  author={Notash, Farzad Y. and He, JiangBiao and Li, Hui},
  journal={IEEE Open J. Power Electron.}, 
  title={Towards Digital Twin Driven High-Reliability Power Electronic Converters}, 
  year={2026},
  volume={7},
  number={},
  pages={683-721},
}

@article{Beiranvand2025,
  title={Computational Power Electronics},
  author={Beiranvand, Hamzeh and Liserre, Marco and Pugliese, Sante and Pascal, Yoann and Langwasser, Marius},
  journal={TechRxiv},
  year={2025, doi: 10.36227/techrxiv.175373769.95905283/v1},
}

@article{Cao2021,
  title={A survey on edge and edge-cloud computing assisted cyber-physical systems},
  author={Cao, Kun and Hu, Shiyan and Shi, Yang and Colombo, Armando Walter and Karnouskos, Stamatis and Li, Xin},
  journal={IEEE Trans. Ind. Inform.},
  volume={17},
  number={11},
  pages={7806--7819},
  year={2021},
  publisher={IEEE}
}

@article{Mazumder2021,
  title={A review of current research trends in power-electronic innovations in cyber-physical systems},
  author={Mazumder, Sudip K and Kulkarni, Abhijit and Sahoo, Subham and Blaabjerg, Frede and Mantooth, H Alan and Balda, Juan Carlos and Zhao, Yue and Ramos-Ruiz, Jorge A and Enjeti, Prasad N and Kumar, PR and others},
  journal={IEEE J. Emerg. Sel. Top. Power Electron.},
  volume={9},
  number={5},
  pages={5146--5163},
  year={2021},
  publisher={IEEE}
}

@article{Yohanandhan2020,
  title={Cyber-physical power system ({CPPS}): A review on modeling, simulation, and analysis with cyber security applications},
  author={Yohanandhan, Rajaa Vikhram and Elavarasan, Rajvikram Madurai and Manoharan, Premkumar and Mihet-Popa, Lucian},
  journal={IEEE Access},
  volume={8},
  pages={151019--151064},
  year={2020},
  publisher={IEEE}
}

@article{Bose2020,
  title={Artificial intelligence techniques: How can it solve problems in power electronics?: An advancing frontier},
  author={Bose, Bimal K},
  journal={IEEE Power Electron. Mag.},
  volume={7},
  number={4},
  pages={19--27},
  year={2020},
  publisher={IEEE}
}

@article{Chen2025,
  title={Power for {AI} and {AI} for power: The infinite entanglement between artificial intelligence and power electronics systems},
  author={Chen, Minjie and Cui, Han and Blaabjerg, Frede and Lorenz, Leo and Hellinger, Rolf and Gray, Tom and Fink, Olga and Hermanns, Kevin},
  journal={IEEE Power Electron. Mag.},
  volume={12},
  number={1},
  pages={37--43},
  year={2025},
  publisher={IEEE}
}

@book{Sarkka2023,
  title={Bayesian Filtering and Smoothing},
  author={S{\"a}rkk{\"a}, Simo and Svensson, Lennart},
  year={2nd ed., 2023},
  publisher={Cambridge University Press}
}

@book{Anderson1979,
  title={Optimal Filtering},
  author={Anderson, Brian DO and Moore, John B},
  year={1979},
  publisher={Prentice-Hall}
}

@article{Hanlon2000,
  title={Multiple-model adaptive estimation using a residual correlation {Kalman} filter bank},
  author={Hanlon, Peter D and Maybeck, Peter S},
  journal={IEEE Trans. Aerosp. Electron. Syst.},
  volume={36},
  number={2},
  pages={393--406},
  year={2000},
  publisher={IEEE}
}

@article{Dingler2022,
  title={State estimation with the Interacting Multiple Model {(IMM)} method},
  author={Dingler, Sebastian},
  journal={arXiv preprint arXiv:2207.04875},
  year={2022}
}

@article{deLopezDiz2023,
  title={A real-time digital twin approach on three-phase power converters applied to condition monitoring},
  author={de Lopez Diz, Sergio and L{\'o}pez, Roberto Mart{\'\i}n and S{\'a}nchez, Francisco Javier Rodr{\'\i}guez and Llerena, Edel D{\'\i}az and Pe{\~n}a, Emilio Jos{\'e} Bueno},
  journal={Appl. Energy},
  volume={334},
  pages={120606},
  year={2023},
  publisher={Elsevier}
}

@article{Liu2024,
  title={One-cycle digital twin-based multiparameter identification of power electronic converters with simple implementation and high accuracy},
  author={Liu, Yisi and Qing, Xinlin and Chen, Guipeng},
  journal={IEEE Trans. Instrum. Meas.},
  volume={73},
  pages={1--11},
  year={2024},
  publisher={IEEE}
}

@article{Kuprat2024,
  title={Thermal digital twin of power electronics modules for online thermal parameter identification},
  author={Kuprat, Johannes and Debbadi, Karthik and Schaumburg, Joscha and Liserre, Marco and Langwasser, Marius},
  journal={IEEE J. Emerg. Sel. Top. Power Electron.},
  volume={12},
  number={1},
  pages={1020--1029},
  year={2024},
  publisher={IEEE}
}

@article{Wang2026b,
  author={Wang, Shaoshuai and Ma, Ruiqing},
  journal={IEEE J. Emerg. Sel. Top. Power Electron.}, 
  title={An Embedded Digital Twin of {DC-DC} based on Hybrid {RLS} for Real-Time Condition Monitoring}, 
  year={(Early Access), 2026},
  volume={},
  number={},
  pages={},
}

@article{BehzadNazif2025,
  author={Behzad Nazif, Parsa and Saeed, Mariam and Ahmad, Saad and Manuel Guerrero, Juan and Rodríguez Mendez, Aitor and Carlos Ozaita Araico, Guillermo and Larrazabal, Igor and Briz, Fernando},
  journal={IEEE J. Emerg. Sel. Top. Power Electron.}, 
  title={Parameter Identification for {DC-DC} Buck Converter Digital Twin Considering Sensor Errors}, 
  year={2025},
  volume={13},
  number={4},
  pages={5111-5123},
}

@article{Wunderlich2023,
  title={Closed-form implicit models for efficient simulation of power electronics},
  author={Wunderlich, Andrew S and Santi, Enrico},
  journal={IEEE J. Emerg. Sel. Top. Power Electron.},
  volume={11},
  number={2},
  pages={1568--1577},
  year={2023},
  publisher={IEEE}
}

@article{Alcaide2024,
  author={Alcaide, Abraham Marquez and Buticchi, Giampaolo and Chub, Andrii and Dalessandro, Luca},
  journal={IEEE J. Emerg. Sel. Top. Ind. Electron.}, 
  title={Design and Control for High-Reliability Power Electronics: State-of-the-Art and Future Trends}, 
  year={2024},
  volume={5},
  number={1},
  pages={50-61},
}

@article{Zhao2022,
  title={Parameter estimation of power electronic converters with physics-informed machine learning},
  author={Zhao, Shuai and Peng, Yingzhou and Zhang, Yi and Wang, Huai},
  journal={IEEE Trans. Power Electron.},
  volume={37},
  number={10},
  pages={11567--11578},
  year={2022},
  publisher={IEEE}
}

@article{Song2024,
  title={Digital twin modeling method of three-phase inverter-driven {PMSM} systems for parameter estimation},
  author={Song, Wensheng and Zou, Yuchao and Ma, Chenwei and Zhang, Sihui},
  journal={IEEE Trans. Power Electron.},
  volume={39},
  number={2},
  pages={2360--2371},
  year={2024},
  publisher={IEEE}
}

@article{Rauf2025,
  title={Novel digital twin deployment approaches: Local and distributed digital twin},
  author={Rauf, Shahid and Muhammad, Fazal and Badshah, Akhtar and Alasmary, Hisham and Waqas, Muhammad and Chen, Sheng},
  journal={IEEE Access},
  volume={13},
  pages={72142--72152},
  year={2025},
  publisher={IEEE}
}

@book{Proakis2008,
  author={Proakis, J. G. and Salehi, M.}, 
  title={Digital Communications},
  publisher={McGraw-Hill},
  year={5th ed., 2008}
}

\end{document}